\documentclass{article}

\usepackage{arxiv}

\usepackage[utf8]{inputenc} % allow utf-8 input
\usepackage[T1]{fontenc}    % use 8-bit T1 fonts
\usepackage{hyperref}       % hyperlinks
\usepackage{url}            % simple URL typesetting
\usepackage{booktabs}       % professional-quality tables
\usepackage{amsfonts}       % blackboard math symbols
\usepackage{nicefrac}       % compact symbols for 1/2, etc.
\usepackage{microtype}      % microtypography
\usepackage{lipsum}		% Can be removed after putting your text content
\usepackage{graphicx}
\usepackage{natbib}
\usepackage{doi}
\usepackage{amsmath}
\usepackage{threeparttable}
\usepackage{multirow}

\title{Evaluation of machine-learning models to measure individualized treatment effects from randomized clinical trial data with time-to-event outcomes}

\author{ \href{https://orcid.org/0000-0001-8014-4431}{\includegraphics[scale=0.06]{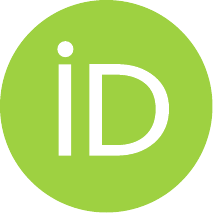}\hspace{1mm}Elvire ~Roblin} \\
	MICS, CentraleSupélec\\
	Oncostat CESP U1018, Inserm,\\
	Université Paris-Saclay, France\\
	\And
	\href{https://orcid.org/0000-0001-7679-6197}{\includegraphics[scale=0.06]{orcid.pdf}\hspace{1mm}Paul-Henry ~Cournède} \\
	MICS, CentraleSupélec,\\
    Université Paris-Saclay,\\
    Gif-sur-Yvette, France 
    	\And
	\href{https://orcid.org/0000-0002-6963-2968}{\includegraphics[scale=0.06]{orcid.pdf}\hspace{1mm}Stefan ~Michiels} \\
	Oncostat CESP U1018, Inserm,\\
     Université Paris-Saclay, \\
    Villejuif, France 
}

\date{}

\renewcommand{\headeright}{}
\renewcommand{\undertitle}{}
\renewcommand{\shorttitle}{Evaluation of ML models to measure ITE in RCT}

\hypersetup{
pdftitle={A template for the arxiv style},
pdfsubject={q-bio.NC, q-bio.QM},
pdfauthor={Elvire ~Roblin, Paul-Henry ~Cournède, Stefan ~Michiels},
pdfkeywords={Randomized Control Trial, Feedforward Neural Network, Interaction Forest, LASSO, Personalized Medicine, Individualized Treatment Effect},
}

\begin{document}
\maketitle

\begin{abstract}
\textbf{Objective:} In randomized clinical trials, prediction models can be used to explore the relationships between patients’ variables and treatment effect magnitude. Our aim was to evaluate flexible machine learning models that incorporate interactions and nonlinear effects from high-dimensional data to estimate individualized treatment recommendations in trials with time-to-event outcomes.\\
\textbf{Methods:} We compared neural networks (CoxCC and CoxTime) and random survival forests, including Interaction Forests (IF) and Causal Survival Forest (CSF) against a Cox proportional hazards model with an adaptive LASSO (ALASSO) penalty as a benchmark. For individualized treatment recommendations, we adapted metrics originally designed for binary outcomes to accommodate time-to-event data with censoring, using both survival-probability and restricted mean survival time estimates of treatment benefit. Metrics included the C-for-Benefit, E50-for-Benefit, Integrated Calibration Index for benefit, and root-mean-squared error for treatment benefit. We also evaluated decision-making utility using the Area Under the Targeting Operator Characteristic and Population Benefit (PB). An extensive simulation study used two data generation processes incorporating nonlinearity and interactions. The models were applied to three cancer clinical trial datasets.\\
\textbf{Results:} 
In the first data generation process, neural networks outperformed ALASSO in calibration while IF showed superior C-for-benefit performance. In the second one, machine learning methods outperformed the benchmark linear ALASSO method across discrimination, calibration, and RMSE. In the cancer trial datasets, the machine learning methods often performed better than ALASSO: neural networks consistently achieved the best calibration, IF achieved the highest C-for-Benefit, and CSF showed strong calibration and the highest PB in the trastuzumab dataset. \\
\textbf{Conclusion:}
Machine learning-based survival models can efficiently estimate conditional average treatment effects from randomized trials when strong nonlinear and interaction effects are expected. These flexible approaches offer valuable alternatives to traditional regression methods for developing treatment recommendation rules from clinical trials with time-to-event outcomes and medium to high-dimensional data set sizes.
\end{abstract}


\section{Introduction}\label{sec1}
Precision medicine involves identifying the intervention that is most likely to be beneficial for a given patient. It is defined by the European Society for Medical Oncology \citep{Yates2018} as a "healthcare approach with the primary aim of identifying which interventions are likely to be of most benefit for which patients based upon the features of the individual and their disease". It focuses on characterizing the variability in patients' response to treatment and estimating treatment benefits. 

In randomized controlled trials (RCTs), in which patients are randomly assigned to two treatment groups: one that receives the experimental treatment, often called the treatment group, and a second group that does not receive the treatment, often called the control group, the treatment benefit is typically calculated as an average across the study population. In an RCT, an individual treatment benefit can be defined as the difference in survival outcomes for a given patient at a pre-specified time point with and without the treatment. The assessment of this treatment benefit is based on a counterfactual approach \citep{Rubin1974}, as both outcomes cannot be observed for a given individual. 

In this work, we explore the potential of two families of machine learning methods to estimate conditional average treatment effects from data in a medium to high-dimensional setting and in the context of an RCT, aiming to predict individualized treatment effects. First, we consider feedforward neural networks (FNNs) adapted to survival analysis through specialized loss functions in a continuous-time framework, namely CoxCC and CoxTime \citep{Kvamme2019}. Second, we study forest-based methods designed to capture treatment effect heterogeneity. These include Interaction Forests (IF), which rely on diversity forests to model bidirectional interactions \citep{Hornung2021}, and Causal Survival Forests (CSF), a generalized random forest approach that directly targets heterogeneity of the conditional treatment effect on the survival scale \citep{Cui2023}. We benchmark these methods against a penalized Cox Proportional Hazards (CoxPH) model \citep{Cox1972} with an adaptive LASSO penalty, as previously recommended by \cite{Ternes2017} for this setting.

Our study examines the potential of these methods to estimate individual patient treatment benefits from data in a high-dimensional setting, where nonlinear and interaction effects are present. We apply calibration and discrimination measures specifically designed for the potential outcome framework and adapt some of these measures to a time-to-event outcome.

 The performances of the different models are evaluated and compared using a simulation study with two different data generation processes, which simulate data with the presence of nonlinearity and interactions. We also present the results of two case studies from clinical trials in cancer utilizing gene expression data. 

\section{Related work}\label{sec2}
In an RCT, the main analysis focuses on the average treatment effect, which refers to the impact of a treatment on a specific outcome variable. This outcome can be of various types, such as a binary outcome or time-to-event. In this work, we are interested in the heterogeneity of the treatment effect based on patient characteristics. An overview of recent developments in statistical methods for assessing the heterogeneity of treatment effects, with a particular focus on subgroup identification and the estimation of individualized treatment strategies is provided in \cite{Lipkovich2024}. For survival outcomes, the FNN model DeepSurv \citep{Katzman2016} provide a personalized recommendation method derived from randomized data, which may predict survival in a subgroup of patients. \cite{Klaveren2017} focus on outcome risk and define treatment benefit as the difference between outcome risk with and without therapy. They define treatment benefit in 2 different ways. Observed treatment benefit is the difference in survival outcome between 2 patients with the same predicted benefit but with different treatment assignments. The predicted treatment benefit is the predicted risk with a treatment for a given patient minus the predicted risk with an alternative treatment option for the same patient. 

In the counterfactual framework, a matching procedure should be implemented to compute the observed treatment benefit. Different methods can create pairs of patients with similar profiles but different treatment assignments. Patients can be matched on predicted benefit \citep{Klaveren2017}: $2$ patients are paired if they have close predicted benefit and discordant treatment assignments. \cite{Klaveren2017} compare this procedure with an alternative matching based on covariates. This second matching procedure is also applied by \cite{Maas2023}. They use the Mahalanobis distance between patients' characteristics to create pairs of patients. 

\cite{Klaveren2017} study binary outcomes and predicted risk, discussing how to extend their measure to time-to-event data. \cite{Rekkas2023} estimate individualized treatment effects from an RCT, comparing various risk-based approaches with a binary outcome. \cite{Bouvier2024} investigates methods for evaluating individualized treatment effects with time-to-event outcomes using individual participant data meta-analysis, highlighting the potential of multiple clinical trial data sets in a meta-analysis to predict individualized treatment effects.  

\section{Material and Methods}\label{sec3}
\subsection{Notations}\label{subsec31}
Let the random variable $T \in \mathbb{R}^{+}$ represent the survival time, that is, the time between the starting point and the occurrence of a given event (e.g., the time between a patient's treatment assignment and death).  The survival function at time $t$, $S(t)$, is defined as:
\begin{equation}
    S(t) = P(T>t). \nonumber
\end{equation}
The survival function can be obtained through the cumulative hazard $H(t)$:
\begin{equation}
    S(t) = \exp(-H(t)), \nonumber
\end{equation}
with $H(t)=\int_{0}^{t}h(s)ds$ , and $h(t)$ the hazard rate. $H_{0}$ is the cumulative hazard at baseline for time $t=0$.

Often, $T$ is not observed for all individuals: time-to-event data is censored. Survival data are said to be right-censored at the last date of follow-up, meaning that no event has occurred. In the case of right-censoring, we define $(C_{i})_{i=1,\dots,n}$ the independent and identically distributed censoring times of $n$ individuals. We can set $\{\tilde{T}_{i},X_{i},D_{i}\}_{i=1,\dots,n}$ where $\tilde{T}=\min(T_{i},C_{i})$ is the time until death or censoring, $X_{i}=(X_{i,1},\dots,X_{i,p})^{T} \in \mathbb{R}^{p}$ denotes a $p$-dimensional vector of variables, and $D_i=\mathbb{I}\{\tilde{T}_{i}=T_{i}\}$ is the censoring indicator. We suppose that the survival time $T_{i}$ is independent of the censoring time conditionally on the vector of variables $X_{i}$ for $i=1,\dots,n$. $G(t)$ can be defined as the estimate of the censoring survival function:
\begin{equation}
   G(t) = P(C>t).  \nonumber
\end{equation}

\subsection{Models}\label{subsec32}

In this paper, we compare $3$ survival-adapted machine learning algorithms and investigate their ability to estimate conditional average treatment effects. 
The treatment effect can be defined as the effect of treatment $\tau$ on the outcome variable $Y$. Since $Y$ is the observed outcome, we have:
\begin{equation}
  Y =
    \begin{cases}
      Y(1) & \text{if $\tau=1$ (treatment)} \nonumber\\
      Y(0) & \text{if $\tau=0$ (control)}. \nonumber
    \end{cases}       
\end{equation}
When $\tau=1$, $Y(0)$ is not observed: it is the counterfactual \citep{Rubin1974}.

In an RCT, assuming that $Y(1)$ and $Y(0)$ are independent from $\tau|X$, the conditional average treatment effect can be defined as: 
\begin{equation}
    \mathbf{E}(Y(1)-Y(0)|X) = \mathbf{E}(Y|\tau=1,X)-\mathbf{E}(Y|\tau=0,X)\nonumber
\end{equation}
with $X$ the characteristics of the patient.

More specifically, we compare $2$ FNNs that use specific loss functions to handle censored observations with IF, a specific type of random survival forest (RSF) that models interaction effects. These models are benchmarked with a penalized CoxPH with linear effects and the LASSO penalty.

\subsubsection{Benchmark method: CoxPH with adaptive LASSO penalty}\label{subsubsec321}
The Cox Proportional Hazards (CoxPH) model \citep{Cox1972} is the most commonly used regression method in survival analysis. In a regression model, the objective is to estimate regression coefficients to assess the strength of association between the predictors $X$ and the outcome. For a given patient $i$, represented by a triplet $(X_{i}, \tilde{T}_{i}, D_{i})$, the hazard function $h(t,X_{i})$ examines how variables influence the rate of a particular event happening at a specific time $t$. It is written as:
\begin{equation}
    h(t|X_{i})=h_{0}(t)\exp(\beta^{T}X_{i}), \text{ for }i=1,\dots,n,
\end{equation}
where the baseline hazard function, $h_{0}(t)$, can be an arbitrary non-negative function of time. $\beta^{T}=(\beta_{1},\dots,\beta_{p})$ is the coefficient vector associated with the variables.

This CoxPH model relies on two assumptions. First, it supposes a linear relationship between the log hazard and the variables $X_{i}$, with the baseline hazard being an intercept term that varies with time. Additionally, the ratio of the instantaneous risk for any $2$ patients is independent of time $t$: this corresponds to the proportional hazard assumption.

With high-dimensional data, the model can become non-identifiable. A LASSO penalty can be introduced to select only the variables with the strongest effects on the outcome of interest. First introduced in the context of linear regression by \cite{Tibshirani1996} and then adapted to survival analysis \citep{Tibshirani1997}, the LASSO  penalty corresponds to the $L_{1}$ norm of the regression coefficients. The penalization term is denoted by: 
\begin{equation}
    pen(\lambda) = \lambda||\beta||_1 \nonumber = \lambda\left( \sum_{j=1}^p |\beta_j| \right). \nonumber
    \label{equ:penlambda}
\end{equation} 

The model's degree of parsimony or complexity depends on the penalization parameter $\lambda$: it varies from $0$ (complete model including all biomarkers) to $+\infty$ (null model with no biomarkers). $\lambda$ is often estimated using cross-validation (CV).

In certain scenarios, LASSO can be inconsistent for variable selection. Here, we implement adaptive lasso (ALASSO) \citep{Fan2008}, a method based on an additive penalization term, which provides a more conservative choice of the shrinkage parameter $\lambda$. With ALASSO, adaptive weights are used for penalizing difference coefficients in the $L_{1}$ penalty.  The penalization term becomes: 
\begin{equation}
    pen(\lambda) = \lambda\left( \sum_{j=1}^p w_j |\beta_j| \right),
\end{equation} 
where $w_j = \frac{1}{|\hat{\beta}_{j}|}$ are biomarker-specific weights, and $\hat{\beta}_{j}, j=1,\dots, p$ are estimated by fitting a preliminary regular LASSO. The non-negative penalty parameter $\lambda$ for the adaptive lasso is chosen using the maximum cross-validated log-likelihood method. 

\subsubsection{Machine learning algorithms adapted to time-to-event outcomes}\label{subsubsec322}

The hypotheses of the linear CoxPH model are restrictive, for instance, when nonlinear and complex interactions exist in the data. Here, we focus on artificial neural networks and random forests, as these algorithms are well-suited to model complex interaction patterns.

\cite{Kvamme2019} introduced a method called CoxCC that uses a loss function based on a case-control approximation. They proposed randomly sampling a new set of controls at each iteration instead of keeping the control samples fixed. The loss is written as:
\begin{equation}
\mathcal{L}_{\text{CoxCC}} = \frac{1}{n}\sum\limits_{i:D_{i}=1}\log(\sum\limits_{j\in \tilde{\mathcal{R}}_{i}}\exp[\phi(X_{j})-\phi(X_{i})]),
\label{equ:loss_coxcc}
\end{equation}
with $\tilde{\mathcal{R}}_{i}$ a subset of the risk set $\mathcal{\mathcal{R}}_{i}$ at time $t$ including individual $i$. $i$ represents the case, and the $j$'s are the controls sampled from the risk set.  $\phi$ represents the transformation operated by the FNN on the input variables.  CoxCC can be trained with this specific loss (Equation~\ref{equ:loss_coxcc})  using a mini-batch gradient descent algorithm.

The authors also developed a second version of their model, CoxTime, that is not constrained by the proportionality assumption. To do so, they added the time variable as an additional input to the model. The loss function can then be rewritten as:

\begin{equation}
    \mathcal{L}_{\text{Cox-Time}} = \frac{1}{n}\sum\limits_{i:D_{i}=1}\log\left(\sum\limits_{j\in \tilde{\mathcal{R}}^{i}}\exp\left[\phi(t_i,X_{j})-\phi(t_i,X_{i})\right]
\right).
    \label{equ:loss_coxtime}
\end{equation}

In this paper, we compare the FNNs with IF \citep{Hornung2021}, a type of random survival forest specifically designed to model quantitative and qualitative interaction effects in bivariate splits. $2$ variables are said to interact if the effect of one variable on the outcome depends on the value of the other variable. Here, we focus on biomarker-by-treatment interactions in the setting of an RCT. Two types of interactions are accounted for,  quantitative and qualitative interactions, as categorized by  \cite{Peto1982}. For a quantitative interaction, the treatment effect varies quantitatively depending on the biomarker value, but remains in the same direction, conditional on the biomarker value (e.g., the treatment effect increases as the biomarker value increases). Conversely, when considering a qualitative interaction, the direction of the treatment effect depends on the value of the biomarker. 

\cite{Cui2023} extend the causal forest framework \citep{Athey2019} to right-censored time-to-event data by introducing the causal survival forest (CSF), a method for estimating potentially heterogeneous CATE under censoring. The approach targets treatment effects defined on transformed survival outcomes, such as the restricted mean survival time (RMST) up to a horizon $t^*$, or differences in survival probabilities at a fixed time point, t. The method proceeds by first estimating nuisance components, including propensity scores, outcome regressions, and the survival and censoring distributions, using flexible machine learning models. These estimates are then incorporated into a doubly robust, Neyman-orthogonal estimating equation, which ensures consistency if either the survival or censoring model is correctly specified. A forest is subsequently trained to directly target CATEs by maximizing heterogeneity in pseudo-outcomes derived from this orthogonal score; causal survival forests split nodes to maximize differences in treatment effects between subgroups, thereby identifying patients who benefit from or are harmed by treatment. Final estimates are obtained by solving a localized weighted estimating equation using forest-derived weights. In randomized trial settings, known treatment assignment probabilities can be incorporated by specifying the propensity score accordingly.

\subsubsection{Hyperparameter search}\label{subsubsec323}

We implemented model-specific hyperparameter tuning strategies tailored to the type of model and dataset (synthetic or real). For the synthetic datasets, the data were first split into training and test sets; hyperparameters were then selected using nested cross-validation on the training set for ALASSO, CoxCC, and CoxTime, while IF and CSF were trained with recommended hyperparameters.
For ALASSO, hyperparameter search was restricted to the regularisation strength $\lambda$, selected automatically by 5-fold cross-validation on the training set. After $\lambda$ selection, variables with non-zero coefficients at the optimal $\lambda$ were retained, and the Cox model was refit on the full training set with those coefficients fixed to obtain survival predictions. The treatment indicator was forced into the model (unpenalised).

For FNN models, we performed hyperparameter optimization with the Tree-structured Parzen Estimator (TPE) algorithm \citep{Bergstra2011}, with 200 trials per simulation. For each trial, the training set was randomly partitioned into K=5 folds; at each iteration the model was trained on K-1 folds and validated on the remaining fold, until each fold had served as the validation set, with early stopping on the held-out fold. The objective minimised by TPE was the minimum across epochs of the mean validation loss across the five folds, where the loss is the Cox partial log-likelihood approximated by case-control sampling. The search space (Supplemental Material Table~\ref{HyperparameterTable}) covered the number of hidden layers, number of neurons per layer, dropout rate, $L_2$ regularisation, learning rate, batch size, optimiser, and activation function. After selection, the model was refit on the training set before evaluation on the held-out test set.

Both forest-based methods were trained with recommended default hyperparameters, as random forests are known to be relatively robust to such choices \citep{Probst2019}. For IF, the number of variables sampled at each split was set to $\sqrt{p}/2$ and each forest consisted of 2,000 trees. For CSF, we used 2,000 trees and $\min\left(\left\lceil\sqrt{p}\right\rceil + 20,\ p\right)$ variables sampled at each split. As we are only in the randomised-trial setting, the propensity was fixed to the known treatment-allocation probability. The forest was trained on the training set and evaluated on the test set.

For the real data analyses, we used a nested double 5-fold cross-validation scheme \cite{Matsui2012} applied to the full dataset to better approximate external validation, given the limited sample sizes. The dataset was partitioned into five folds, and each fold was held out in turn as a test set (outer loop). For each outer split, the remaining data were used for model training and hyperparameter tuning via an inner 5-fold cross-validation. Hyperparameter optimisation within the inner loop was performed for ALASSO, CoxCC, and CoxTime using the same procedures as in the synthetic data setting; IF and CSF were trained with the same default hyperparameters described above, without inner-loop tuning. For the tuned models, the configuration minimizing the average validation loss across inner folds was selected. The model was then refit on the outer training folds using the selected hyperparameters (when applicable) and evaluated on the corresponding outer test fold. This process was repeated across all outer folds. To further reduce variability in performance estimates, the entire nested cross-validation procedure was repeated over $B$ bootstrap resamples of the dataset.

\subsection{Data}\label{subsec33}

The models are trained on two types of data: simulated data and real patient data sets. The synthetic data are generated to examine the operating characteristics and performances of the modeling techniques, while two high-dimensional patient clinical trial data sets are used for illustrative purposes.

\subsubsection{Simulation study}\label{sec331}
 In the first setting, we simulate nonlinear interactions between the biomarkers and the treatment variable using a full biomarker-by-treatment interaction model. 
 
\cite{Rothwell2005} put forward that the only reliable approach for assessing the predictiveness of biomarkers is to test their interaction with the treatment and introduce a full biomarker-by-treatment interaction model. Let $\tau$ denote the treatment indicator, with $\tau=1$ for patients receiving the treatment and $\tau=0$ for those in the control group, and let $X=(X_{1},\dots,X_{p})^{T} \in \mathbb{R}^{p}$ denote the vector of patient covariates. We can simulate data based on this model:   
\begin{equation}
     h(t|\tau,X) = h_{0}(t) \exp\Bigg( \alpha_{\tau}\tau + \sum_{i=1}^{p} \beta_{i1}f_{i}(X_{i})+\sum_{i=1}^{p}\beta_{i2}f_{i}(X_{i})\tau\Bigg).
\end{equation}
In this model, we introduce nonlinearity through the choice of the function $f$, which enables the construction of complex interactions with the treatment. Our model extends the model from \cite{Haller2019}, which considered the case of a single biomarker. \\
Here,  $\alpha_{\tau}$ is chosen as $\alpha_{\tau}=ln(0.75)\approx-0.288$. We set $\beta_{i1} = 0.288$, $f_{i}(X) = (2X-1)^2$ and $\beta_{i2}=-0.9$.   The biomarkers are generated from a uniform distribution: $X_{i} \sim \mathcal{U}_{[0,1]}$. Since we intend to simulate an RCT, a total of $n$ patients per data set are randomly assigned with a 1:1 ratio to the experimental or control arm, with $P(\tau=1)= P(\tau=0) = \frac{1}{2}$.

Survival times are generated from a Weibull distribution. Thus, the risk can be written as follows: 

\begin{equation}
h(t|\tau,X) = b^{-a} a t^{a-1} \exp \Bigg( 
    \alpha_{\tau} \tau 
    + \sum_{i=1}^{p} \beta_{i1} f_i(X_i) 
    \quad + \sum_{i=1}^{p} \beta_{i2} f_i(X_i) \tau 
\Bigg)
\end{equation}

with $a$ the shape parameter and $b$ the scale parameter. For all scenarios, we generate exponential survival times (corresponding to a shape parameter $a=1$) with a time-constant baseline hazard rate. 
Using the inverse of the cumulative risk function for a Weibull distribution, we compute the theoretical survival probabilities (see details in Supplemental material): 
\begin{equation}
    S(t|X) = \exp(-\int_{0}^{t}h(s|X)ds = \exp(-H_{0}(t)\exp(\beta X)) 
\end{equation}

 A cohort of $n$ individuals ($n \in \{2,000; 20,000\}$) is generated for each set, split into a $50\%$ training set and a $50\%$ test set. We generate independent censoring, and consider two censoring rates: a moderate censoring rate of $20\%$ and a high censoring rate of $50\%$. $q=20$ variables are generated for each set. Among the $q$ biomarkers, only $p=10$ are truly prognostic. For each scenario, $100$ replications are performed.

In the second data generation process, nonlinear interactions are simulated using Friedman's random function generator \citep{Friedman2001} within an accelerated failure time (AFT) model. Friedman's random function generator enables us to generate random functions with second-order interactions between the survival data and the explanatory variables, as well as strong nonlinear effects. It is based on an AFT model that assumes a linear relationship between the logarithm of survival time $T$ and the variables. The random function generator is defined as: 
 \begin{equation}
     log\ T = m(X) + W  \text{ with } W \sim \Gamma(2,1). \nonumber
 \end{equation}
 
A vector of variables $X = (X_{1},...,X_{20})$ is generated with $X \sim \mathcal{N}(0,I)$. $m(X)$ is defined by Friedman's random function generator: 
\begin{equation}
 m(X) = \sum_{l=1}^{10} a_{l}g_{l}(R_{l}). \nonumber
\end{equation}
 
$\{a_{l}\}_{1}^{10}$ are randomly generated from a uniform distribution ($a_{l} \sim\mathcal{U}_{[-1,1]}$). $R_{l}$ is a random subset of the input vector $X$ of size $n_l=5$, implying high-order interaction effects. 
With Friedman's function generator, the input variables are associated with the survival time at different levels: 
\begin{equation}
g_{l}(R_{l}) = \exp \{ -\frac{1}{2}(R_{l}-\mu_{l})^{T}V_{l}(R_{l}-\mu_{l})\}.
\end{equation}
Each mean vector $\{\mu_{l}\}_{1}^{10}$ is randomly generated with $\mu_{l}\sim \mathcal{N}(0,1)$. The matrix of variance-covariance $V_{l}$ is also randomly generated: 
$V_{l}=U_{l}D_{l}U_{l}^{T}$ with $U_{l}$ an orthonormal random matrix, $D_{l} = diag\{d_{l,1},...,d_{l,n_{l}}\}$ and $\sqrt{d_{lk}}\sim \mathcal{U}(0.1,2.0)$. 
Finally, we obtain the survival times by applying the exponential: $\exp(\log(T))$. This is the AFT-Friedman method.
\\
\cite{Henderson2020} adapt the flexible random function generator specifically to treatment effects by introducing a second generator interacting with the treatment variable. For this purpose, the AFT-Friedman model can be rewritten as: 
\begin{equation}
    \log T = m_{1}(X) + \tau\times m_{2}(X) + W \text{ with } W\sim \Gamma(2,1), 
\end{equation}
where $m_{2}(X)$ is defined as : 
\begin{equation}
  m_{2}(X) = \sum\limits_{i=1}^{10}a_{2l}g_{2l}(R_{2l}). \nonumber
\end{equation}
 The coefficients $a_{2l}$  are randomly generated from a uniform distribution $a_{2l}\sim \mathcal{U}_{[-0.2,0.3]}$. A total of $n$ patients per data set were randomly assigned $(1:1)$ to the experimental or control arm, with $P(\tau=1)= P(\tau=0) = \frac{1}{2}$. 

By using the inverse of the cumulative risk function for a log-normal distribution, we are able to compute the theoretical survival probabilities (details in Supplemental material) :  
\begin{align}
    S(t|X) &= \exp \Bigg(-\int_{0}^{t}h(s|X)ds\Bigg) \nonumber  \\
    &= \exp\Bigg(-H_{0}(t\exp(m_{1}(X) + \tau\times m_{2}(X))\Bigg) 
\end{align}

 A cohort of $n$ individuals ($n \in \{2,000; 20,000\}$) is generated for each set, split into $50\%$ for the training set and $50\%$ for the test set. We generate independent censoring, and consider a moderate censoring rate of $20\%$. For each scenario, $100$ replications are performed.

\subsection{Cases studies: application on cancer data sets}\label{subsec34}

We  apply the methods to gene expression data from a meta-analysis of neoadjuvant trials that included $614$ breast cancer patients treated by anthracyclines alone or anthracyclines plus taxanes \citep{Ternes2018}. We also use gene expression data from an RCT including $1,574$ patients, which evaluated the effect of adjuvant trastuzumab in early breast cancer \citep{PogueGeile2013}, and finally gene signature data an RCT in renal cancer with $n = 886$ patients,  which assessed first-line avelumab plus axitinib versus sunitinib in previously untreated patients with advanced renal cell carcinoma \citep{Motzer2020}.

\subsubsection{Breast cancer study on taxane chemotherapy}\label{subsubsec341}
We apply the models to  a meta-analysis of neoadjuvant trials that included $614$ breast cancer patients treated by anthracycline-based chemotherapy with ($n=507$) or without ($n=107)$ taxane. The dataset comprises $p=1,689$ genes. This cohort is available in the  biospear \citep{Ternes2018} R package, which is publicly available. 

\begin{figure*}[!ht]
\centering
\includegraphics[width=0.4\textwidth]{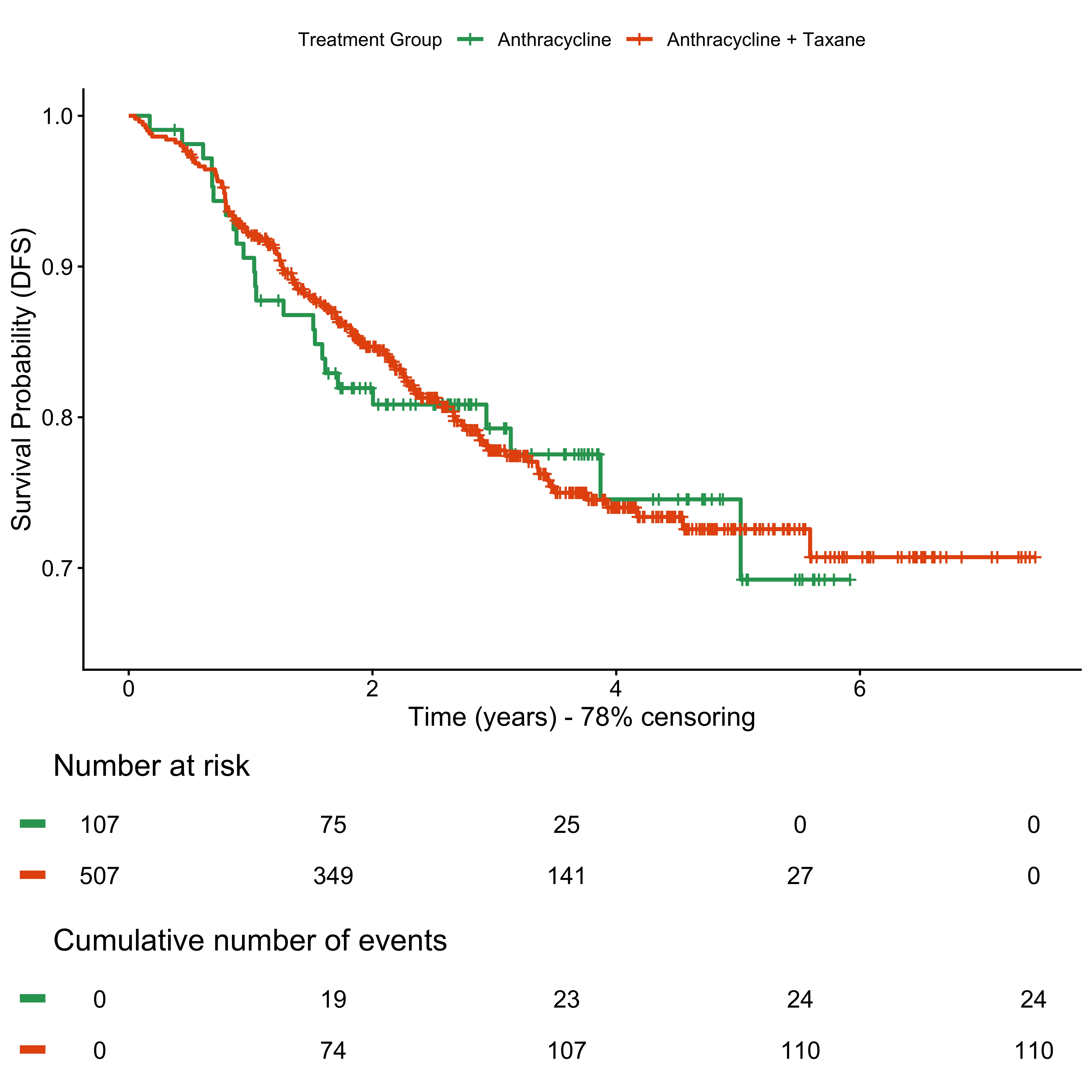}
\caption{Kaplan-Meier curves for the breast cancer cohort on taxane chemotherapy.}\label{fig:KMBreast}
\end{figure*}

\subsubsection{Breast cancer cohort with adjuvant trastuzumab}\label{subsubsec342}
The models are also compared on data from an RCT funded by the National Cancer Institute and the National Institute of Health. This randomized phase III clinical trial evaluated the effect of adding trastuzumab to adjuvant chemotherapy on disease-free survival in early breast cancer patients with HER2$^ {+}$ tumors. \\
 $n = 1,574$ patients were included in the trial: $n = 795$ patients received chemotherapy, while $n = 779$ others received chemotherapy and trastuzumab. For each patient, the expression of $462$ genes is available, in addition to standard clinicopathological variables such as ER status or the number of positive nodes.

\begin{figure*}[!ht]
\centering
\includegraphics[width=0.4\textwidth]{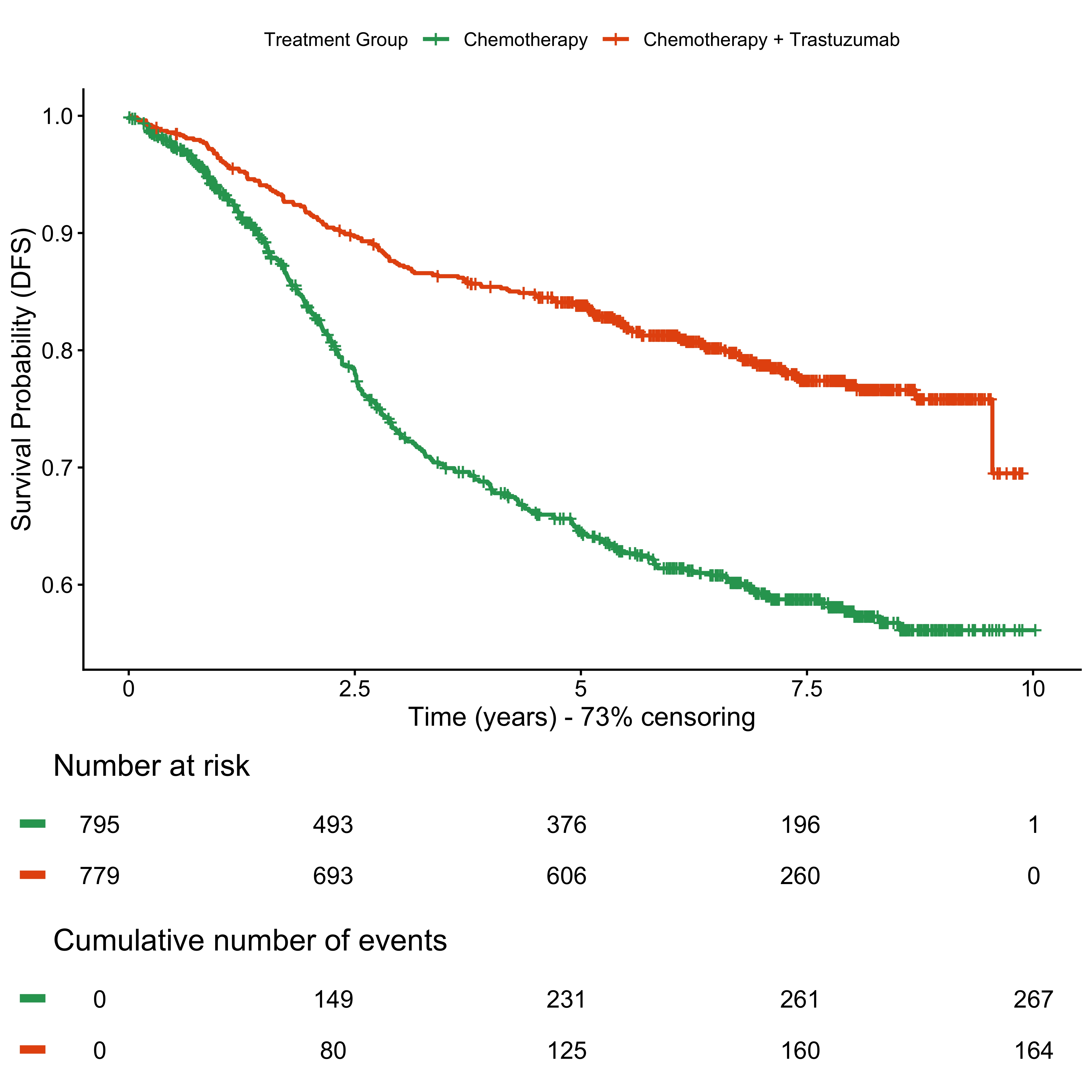}
\label{fig:KMTrastu}
\caption{Kaplan-Meier curves for the breast cancer cohort with adjuvant trastuzumab.}
\end{figure*}

\subsubsection{Advanced Renal Cell Carcinoma cohort with avelumab}
The third cohort analyzed comprises baseline tumor samples from the phase 3 JAVELIN Renal 101 trial, which assessed first-line avelumab plus axitinib versus sunitinib alone for progression-free survival in previously untreated patients with advanced renal cell carcinoma \citep{Motzer2020}. A total of $n = 886$ patients were enrolled, with $n = 443$ assigned to avelumab plus axitinib and $n = 443$ to sunitinib. Consistent with the original study, we utilized the provided pathway scores, where each pathway score for a given sample was computed as the average of the standardized expression values of the genes comprising that pathway. The dataset features $58$ pathway scores alongside $2$ clinical variables: age and sex. 

\begin{figure*}[!ht]
\centering
\includegraphics[width=0.4\textwidth]{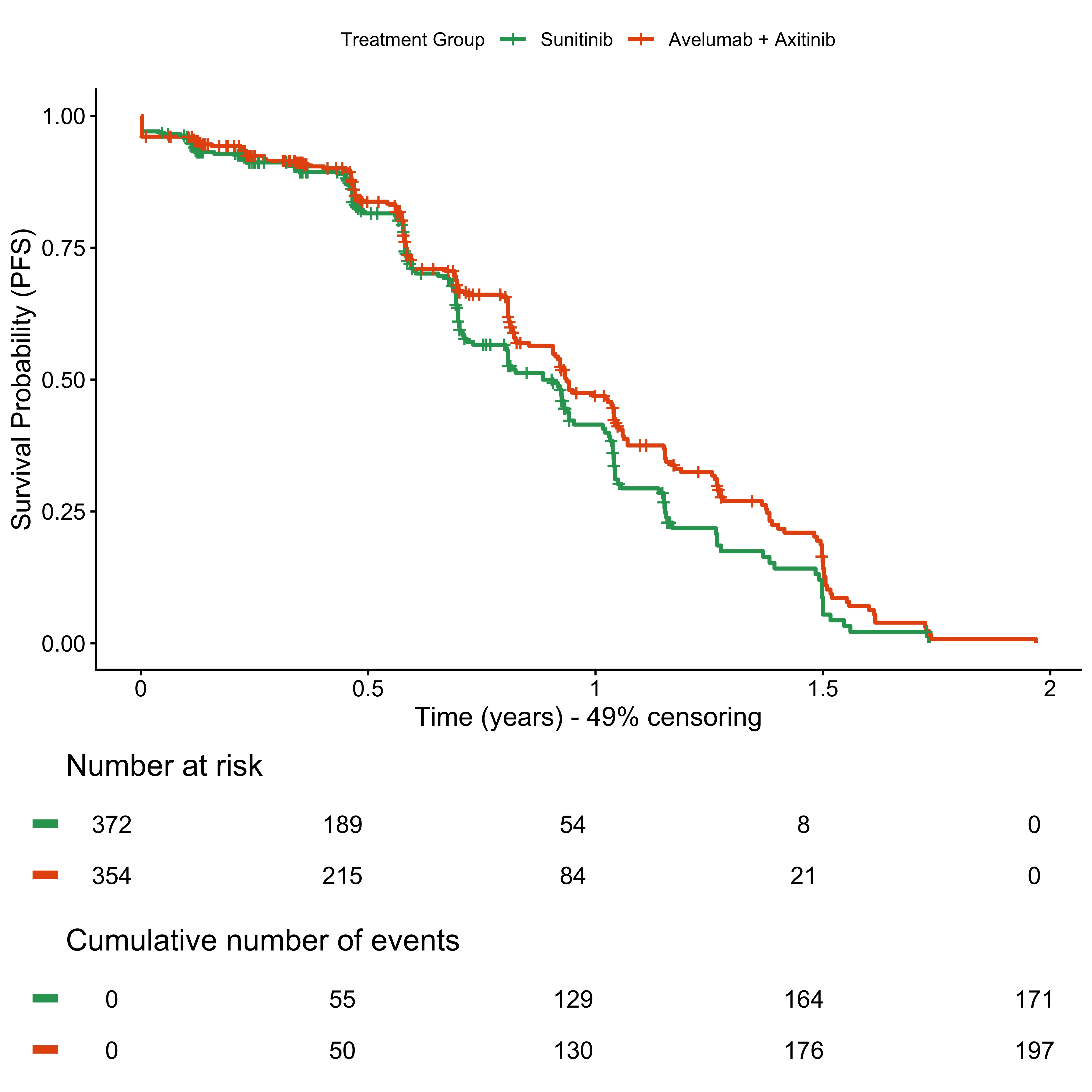}
\label{fig:KMTrastu}
\caption{Kaplan–Meier survival curves for the advanced renal cell carcinoma cohort from the phase III JAVELIN Renal 101 trial.}
\end{figure*}

\subsection{Evaluation Measures}\label{subsec35}

The models are evaluated across four complementary criteria (discrimination, calibration, prediction error, and decision-making utility) using metrics specifically adapted to the assessment of treatment benefit. Discrimination is assessed using the Concordance index for benefit ($\text{C}_{\text{benefit}}$). Calibration is evaluated through the Integrated Calibration Index for benefit ($\text{ICI}_{\text{benefit}}$), and the median calibration error for benefit ($\text{E50}_{\text{benefit}}$). Prediction accuracy is measured using the root-mean-squared error (RMSE), also referred to as the Precision in Estimation of Heterogeneous Effects (PEHE). Finally, decision-making utility is quantified using the Area Under the Targeting Operator Characteristic (AUTOC) and the Population Benefit (PB).

\subsubsection{Definition of treatment benefit}\label{subsubsec351}
We define the conditional average treatment benefit for patient $i$ at time $t$, $\hat{\theta}_i^{SP}$  as an estimate of the CATE at time $t$, computed as the difference between the individual survival probability at a given time point with the treatment option being tested minus the probability with the control for the patient with same covariate values. It is given by:
$$\hat{\theta}_i^{SP}=[\hat{S}(t|\tau=1|X_{i})-\hat{S}(t|\tau=0|X_{i})],$$
where $\hat{S}(t|\tau=a|X_{i})$ denotes the predicted survival probability at time $t$ under treatment assignment $\tau=$, for $a \in \{0,1\}$. To capture treatment benefit over a time interval rather than a single time point, we also define benefit in terms of the RMST up to a clinically relevant horizon $t^{*}$. This quantity is the difference between the expected survival time accumulated up to $t^{*}$ under treatment and under control:
$$\hat{\theta}^{RMST}(t^{*})=\int^{t^{*}}_{0}[\hat{S}(u|\tau=1, X_{i})-\hat{S}(u|\tau=0, X_{i})]du.$$
This formulation has the advantage of preserving the magnitude of benefit over time, rather than reducing it to a single survival probability at one horizon.

\subsubsection{Pairwise observed and predicted treatment benefit}\label{subsubsec352}
Since the counterfactual response for a given patient cannot be observed, the matched patient serves as a surrogate for this unobserved outcome. To construct such pairs, we use two matching procedures, based either on patients’ baseline covariates or on their predicted treatment benefit. For the first method, we compute a predicted benefit score for each individual. Then, within each arm (control and treatment), patients are ranked according to this score, and pairs are formed by aligning the two ranked lists so that patients with similar predicted benefit are matched. For the second method, we apply a covariate-based matching procedure based on patients’ characteristics and the Mahalanobis distance. For each treated patient $u$, we search among all  control patients $v$ and identify the one whose covariate vector $X_{u}$ is closest to $X_{v}$ in terms of Mahalanobis distance. We first compare the two matching strategies on the simulation data, and subsequently retain and apply the predicted-benefit–based matching approach in the real-data analysis.

Given that matching is not perfect, the predicted benefit assigned to a matched pair is defined as the average of the individual predicted benefits of the two patients in the pair. For the survival-probability-based benefit at time t, this gives:
$$\hat{\theta}_{u,v}^{SP}(t)=\frac{\hat{\theta}_{u}^{SP}(t)+\hat{\theta}_{v}^{SP}(t)}{2},$$
with $u$ the patient in the treated arm $(\tau=1)$, and $v$ the patient in the control arm $(\tau=0)$ for a given time $t$. Analogously, the pairwise predicted benefit based on the RMST up to horizon $t^{*}$ is defined as:
$$\hat{\theta}_{u,v}^{RMST}(t^{*})=\frac{\hat{\theta}_{u}^{RMST}(t^{*})+\hat{\theta}_{v}^{RMST}(t^{*})}{2},$$

Observed treatment benefit $\tilde{\theta}$ is defined as the difference in time-to-event for $2$ matched patients. For each pair of patients ($u$ belonging to the control arm, and $v$ to the experimental arm), we calculate the observed treatment benefit using the following formula :  
\begin{equation}
  \tilde{\theta}_{u,v} = \mathbf{1}_{Statut_{u}}\times \mathbf{1}_{\tilde{T}_{u}< \tilde{T}_{v}} - \mathbf{1}_{Statut_{v}}\times \mathbf{1}_{\tilde{T}_{v}< \tilde{T}_{u}}  
\end{equation}

Thus, we have:
\begin{equation}
\tilde{\theta}_{u,v} = 
\begin{cases}
1 & \text{if } D_u = D_v = 1 \text{ and } \tilde{T}_u < \tilde{T}_v, \\
-1 & \text{if } D_u = D_v = 1 \text{ and } \tilde{T}_v < \tilde{T}_u, \\
0 & \text{otherwise.}
\end{cases}
\end{equation}

Specifically, $\tilde{\theta}_{u,v}=1$ if the event time is shorter in one patient, $\tilde{\theta}_{u,v}=-1$ if it is shorter in the other patient, and $\tilde{\theta}_{u,v}=0$ when the ordering is not observable because of censoring or when the two times are tied. This proxy is used for ranking-based evaluation.

\subsubsection{Discrimination measures}\label{subsubsec353}

The C-statistic at time $t$ measures the discrimination ability of a model, indicating its ability to distinguish high-risk and low-risk patients for a fixed time horizon. Specifically, it estimates the probability of agreement, i.e., the probability that $2$ patients randomly selected are ordered similarly in terms of survival prediction and in their observed survival data. Here, we use a concordance measure that accounts for the censored data using the inverse probability of censoring weighting \citep{Uno2011}. The concordance index for a horizon time $t$ is then:
\begin{equation}
\hat{\text{C}}(t)=\frac{\sum\limits_{i=1}^{n}\sum\limits_{j=1}^{n}D_i\hat{G}(\tilde{T}_i)^{-2}\mathbb{I}\{\tilde{T}_i<\tilde{T}_j, \tilde{T}_i<t\}\mathbb{I}\{\hat{S}(t|X_i)<\hat{S}(t|X_j)\}}{\sum\limits_{i=1}^{n}\sum\limits_{j=1}^{n}D_i\hat{G}(\tilde{T}_i)^{-2}\mathbb{I}\{\tilde{T}_i<\tilde{T}_j, \tilde{T}_i<t\}}
\end{equation}

The value of the C-statistic lies between $0$ and $1$, with $0.5$ equivalent to a random prediction and $1$ corresponding to a perfect ability to rank. We compute the C-statistic for all individuals. 

C-for-benefit ($\text{C}_{\text{benefit}}$) was introduced by \cite{Klaveren2017} as an extension of the C-statistic to measure treatment benefit at a horizon time $t$. It represents the probability that, among $2$ randomly chosen matched patient pairs, the pair with greater observed benefit also has a higher mean predicted benefit. It can be written as:
\begin{equation}
    \text{C}_{\text{benefit}}(t) = P(\hat{\theta}_{p_{k}}(t)>\hat{\theta}_{p_{l}}(t)|\tilde{\theta}_{p_{k}}(t)>\tilde{\theta}_{p_{l}}(t)),
\end{equation}
where $p_{k}$ and $p_{l}$ are $2$ pairs of patients.

\subsubsection{Calibration measures}\label{subsubsec354}
To evaluate the calibration capacity of the model, we use measures based on the calibration curve. Calibration commonly refers to the agreement between predicted and observed probabilities for the outcome \citep{Austin2020}. Here, we are interested in the correspondence between the predicted and observed treatment effects.

Calibration can be assessed by a smoothed calibration curve obtained through local regression of the observed pairwise treatment effect on the predicted pairwise treatment effect. The observed benefits are regressed on the predicted benefits using a locally weighted scatterplot smoother (loess). It is a non-parametric regression technique used to create a smoothing function by fitting a curve through a scatterplot of data points. It combines multiple local regressions to fit a curve to subsets of the data, allowing it to capture complex patterns without assuming a specific global model. 

The Integrated Calibration-for-Benefit ($\text{ICI}_{\text{benefit}})$ \citep{Rekkas2023, Maas2023} measures the average difference between predicted and smooth observed benefit across different levels of prediction determined by the empirical density function of the predicted pairwise treatment effect. $\text{E50}_{\text{benefit}}$ represents the $50^{th}$ percentile of the absolute distance between predicted and smooth observed benefit. It assesses how well the predicted treatment effect matches the observed treatment effect at the median. It offers insight into the model’s calibration at the central point rather than across the entire prediction range. For these measures, null values indicate perfect calibration, meaning that the model’s predicted treatment benefits align perfectly with the observed benefits.

The calibration quantile plot \citep{Bouvier2024} represents the average observed benefit versus predicted benefit in quantiles of predicted benefit. The predicted benefit is divided into $5$ bins. In each bin, the average predicted benefit is compared to the observed benefit.

\subsubsection{Prediction errors in simulation}\label{subsubsec355}
The predictive accuracy of the models is assessed using the root mean squared error (RMSE) \citep{Rekkas2023} adapted to time-to-event data, equivalent to the precision in estimation of heterogeneous effects (PEHE) used in the causal-inference literature \citep{Hill2011}. Because the RMSE (PEHE) requires knowledge of the true individual treatment benefit, it can only be computed on simulated data, where the underlying counterfactual survival functions (the true $\theta$) are known by construction. It is defined as $\sqrt{\frac{1}{n} \sum_{i=1}^{n} \left( \theta_i(t) - \hat{\theta}_i(t) \right)^2 }$, where $\theta_i(t)$ denotes the true treatment benefit for individual $i$ at time $t$, and $\hat{\theta}_i(t)$ the corresponding predicted benefit. The RMSE thus quantifies the average deviation between predicted and true individual benefits at the pre-specified horizon $t$, with lower values indicating better predictive accuracy. The metric does not require matching of treated and control patients, because the per-patient counterfactual is known by simulation. It can be computed for both $\hat{\theta}_i^{\mathrm{SP}}(t)$ and $\hat{\theta}_i^{\mathrm{RMST}}(t^*)$, providing complementary views of predictive error at a fixed time point and over the full interval $[0, t^*]$.

\subsubsection{Decision making utility}\label{subsubsec356}
We also assess whether the predicted treatment benefit can be used to support treatment decisions. This is done from two complementary perspectives: prioritization, namely whether the model ranks patients in a way that identifies those who benefit most form treatment, and decision accuracy, namely the population-level survival impact of using the model to decide whom to treat. We address prioritization with the Area Under the Targeting Operator Characteristic (AUTOC), which belongs to the Rank-weighted Average Treatment Effect (RATE) framework of \cite{Yadlowsky2025}, and the decision accuracy with the Population Benefit framework of \cite{Efthimiou2025}.

The RATE framework quantifies the extent to which predicted treatment benefit correctly orders patients according to their observed treatment benefit. In the study, we focus on the Area Under the Targeting Operator Characteristic (AUTOC), a specific RATE estimand obtained by integrating the Targeting Operator Characteristic (TOC) curve over $q \in [0,1]$. For a given fraction $q$, the TOC compares the average treatment effect (ATE) among the top $q$-fraction of patients ranked by predicted benefit with the ATE in the overall population. Using the survival-probability-based predicted benefit at horizon $t$:
\begin{equation}
\mathrm{TOC}^{\mathrm{SP}}(q,t) = \mathrm{ATE}^{\mathrm{SP}}\!\left(t \mid \hat{\theta}_i^{\mathrm{SP}}(t) \in \text{top } q\right) - \mathrm{ATE}^{\mathrm{SP}}(t),
\end{equation}
where $\mathrm{ATE}^{\mathrm{SP}}(t) = \mathbb{E}\!\left[ S(t \mid \tau=1, X) - S(t \mid \tau=0, X) \right]$ is the marginal ATE on the survival-probability scale. The RMST-based version follows by replacing the survival probability with its restricted mean counterpart:
\begin{equation}
\mathrm{TOC}^{\mathrm{RMST}}(q,t^*) = \mathrm{ATE}^{\mathrm{RMST}}\!\left(t^* \mid \hat{\theta}_i^{\mathrm{RMST}}(t^*) \in \text{top } q\right) - \mathrm{ATE}^{\mathrm{RMST}}(t^*),
\end{equation}
where $\mathrm{ATE}^{\mathrm{RMST}}(t^*) = \mathbb{E}\!\left[ \int_0^{t^*} \left( S(s \mid \tau=1, X) - S(s \mid \tau=0, X) \right) ds \right]$.

A predicted benefit that successfully identifies high-benefit patients yields a TOC that is high for small $q$ and decreases toward zero as $q \to 1$. The AUTOC summarizes the curve into a single scalar:
\begin{equation}
\mathrm{AUTOC} = \int_0^1 \mathrm{TOC}(q)\, dq.
\end{equation}
An AUTOC close to zero indicates that the predicted benefit does not separate patients by treatment effect better than chance, whereas an AUTOC above $0$ indicates that the model is informative for identifying patients more likely to benefit from treatment.

While AUTOC evaluates the model's ability to rank patients, the Population Benefit (PB) framework quantifies the survival impact of using the model to decide whom to treat. The model-induced decision rule recommends treatment whenever the predicted benefit exceeds a threshold $\epsilon$, which we set to $0$ for simplicity. Patients are then stratified into four groups according to the sign of their predicted benefit and the treatment received $\tau_i \in \{0,1\}$. $G_1$ contains patients for whom treatment was recommended and received, $G_2$ those for whom it was recommended but not received, $G_3$ those for whom it was not recommended but received, and $G_4$ those for whom it was not recommended and not received. Because treatment is randomized, the comparison $G_1$ vs $G_2$ and $G_3$ vs $G_4$ are unconfounded. The model-based decision rule is therefore evaluated by contrasting outcomes in $G_1 \cup G_4$ (patients who received their model-recommended treatment) against a reference group.

We consider two reference strategies, which answer distinct questions. The model-versus-opposite contrast compares model-concordant assignment to its systematic reversal ($G_2 \cup G_3$). On the survival-probability scale at horizon $t$:
\begin{equation}
\mathrm{PB}^{\mathrm{SP}}_{M\,\text{vs}\,\overline{M}}(t) = \mathbb{E}\!\left[ S(t \mid \text{rule } M, X_i) \right] - \mathbb{E}\!\left[ S(t \mid \text{rule } \overline{M}, X_i) \right],
\end{equation}
where $M$ denotes the model-induced decision rule and $\overline{M}$ its reversal. This estimand tests whether the model is informative at all: if it correctly separates patients who benefit from treatment from those who do not, then reversing its recommendations should yield worse outcomes. In estimation, it is obtained by contrasting survival in $G_1 \cup G_4$ with survival in $G_2 \cup G_3$.

The model-versus-treat-everyone contrast compares model-concordant assignment to a uniform active-treatment policy:
\begin{equation}
\mathrm{PB}^{\mathrm{SP}}_{M\,\text{vs}\,\mathrm{Trt}}(t) = \mathbb{E}\!\left[ S(t \mid \text{rule } M, X_i) \right] - \mathbb{E}\!\left[ S(t \mid \tau_i=1, X_i) \right].
\end{equation}
This estimand is more clinically pragmatic: it asks whether personalized treatment assignment outperforms the simple default of treating all patients, evaluating whether the model adds value beyond a one-size-fits-all policy. In estimation, it is obtained by contrasting survival in $G_1 \cup G_4$ with survival in the subset of patients who received the active treatment ($G_1 \cup G_3$).

On the RMST scale, the same contrasts are obtained by replacing the survival probability $S(t \mid \cdot)$ with the restricted mean survival time up to horizon $t^*$:
\begin{equation}
\mathrm{PB}^{\mathrm{RMST}}_{M\,\text{vs}\,\overline{M}}(t^*) = \mathbb{E}\!\left[ \int_0^{t^*} S(s \mid \text{rule } M, X_i)\, ds \right] - \mathbb{E}\!\left[ \int_0^{t^*} S(s \mid \text{rule } \overline{M}, X_i)\, ds \right],
\end{equation}
and analogously for $\mathrm{PB}^{\mathrm{RMST}}_{M\,\text{vs}\,\mathrm{Trt}}(t^*)$. The RMST formulation summarizes treatment benefit over the full interval $[0,t^*]$ rather than at a single time point, providing a complementary view.

\subsubsection{Variable Importance}\label{subsubsec357}
Beyond predictive performance, we examine which baseline covariates drive the predicted treatment benefit in each model on the 3 cancer data sets. The model families differ in structure, and we therefore use a model-specific importance measure for each: regression coefficients for the ALASSO, SHapley Additive exPlanations (SHAP) values for the neural network models, the Effect Importance Measure (EIM) for the IF, and the built-in variable importance for the CSF. In all four cases, importance is computed per fold, averaged across the 5 cross-validation folds within each bootstrap, and then summarized across the 100 bootstraps. We focus on predictive importance, which reflects how much the conditional average treatment effect varies with $X_j$.

For ALASSO, importance is given by the model coefficients. The model is fitted with explicit treatment-covariate interaction terms $\tau×X_j$, where non-zero coefficient indicates that the variable contributes to the prediction. We focus specifically on interaction effects, quantified by $|\hat{\beta}_{\tau×X_j}|$, to identify predictive biomarkers. For each term, we report the selection frequency (i.e. the proportion of folds in which the coefficient is non-zero) and the mean value of the coefficient when selected.

For the neural network models, the model parameters are not directly interpretable, so we use SHAP values \citep{Lundberg2017} to quantify variable importance. SHAP attributes the deviation of a prediction from its expected value to each input feature. We compute the mean absolute SHAP value of the predicted treatment benefit $\hat{S}(t \mid X,\tau=1)-\hat{S}(t \mid X,\tau=0)$ at pre-specified clinically relevant horizons, yielding predictive importance. When the number of features is large, we restrict reporting to those whose importance lies above the elbow of the sorted importance curve, identified with the kneedle algorithm.

For the IF, we use the EIM, which quantifies the contribution of each variable pair to treatment effect heterogeneity. It allows the ordering of pairs according to the importance of their interaction effects on prediction. It produces three separate rankings of variable pairs and individual variables (univariable, quantitative-interaction, and qualitative-interaction importances) by adapting \cite{Breiman2001} permutation-based variable importance to the univariable and bivariable split types generated by IF. For each tree, the out-of-bag (OOB) prediction accuracy is compared with the accuracy expected if a given variable or variable pair were unavailable: OOB observations are passed through the tree and, at any split involving the target variable (or pair) and matching the targeted split type, are routed randomly to a child node with probability proportional to child-node size. Restricting randomization to splits of the relevant type yields importance values specific to qualitative or quantitative interactions rather than being confounded with marginal effects. For datasets with more than 100 biomarkers, only 5,000 variable pairs are pre-selected, and tree depth is restricted when the dataset includes more than 1,000 patients.

For the CSF, covariates are ranked by their contribution to treatment-effect heterogeneity. The importance of feature $j$ is a depth-weighted count of how often the forest splits on that feature: splits at depth d contribute with weight $(1+d)^{-2}$, only splits up to depth 4 are counted, and the resulting scores are normalized to sum to one across features. Because CSF select splits that maximize heterogeneity in the estimated conditional treatment effect, the splits being counted reflect partitions of patients with different treatment responses, so the importance ranks features by their contribution to treatment-effect heterogeneity.

The importance measures across the different model families are not expressed on a common scale, so we compare rankings rather than raw values. For each model and each notion of importance, we report the selection frequency. $P_{top5}$  (proportion of bootstraps × folds in which the variable appears within the top-5), and the mean rank across all bootstraps and folds. We identify variables that appear consistently among the most important across the different model families as a robust consensus signal of treatment-effect modifiers.

\section{Results}\label{sec4}

\subsection{Results on the simulation sets}\label{subsec41}
\subsubsection{Data generation process 1}\label{subsubsec412}
The results reported here are the ones obtained using data simulated from the Weibull biomarker-by-treatment model with nonlinear interactions.

In Table~\ref{tab:Sett1P5C501}, the highest average $\text{C}_{\text{benefit}}$  was obtained with the IF model, which was also the closest to the oracle value. ALASSO obtained the lowest $\text{C}_{\text{benefit}}$, which could indicate that the linear ALASSO model has difficulty capturing treatment-by-biomarker interactions, whereas IF, due to its specific design for such interactions, demonstrates superior performance in modeling them. Notably, CoxCC and CoxTime yielded similar results, as expected since no interaction with time was generated in the simulation setup. $\text{E50}_{\text{benefit}}$  and $\text{ICI}_{\text{benefit}}$  values remained substantially above zero across all models, highlighting difficulties in obtaining solid calibration or limitations of the matching procedure between control and treatment group patients.  CSF displayed markedly lower $\text{E50}_{\text{benefit}}$  and $\text{ICI}_{\text{benefit}}$  values compared to all other models. 

Among machine learning methods, IF and CSF performed well in terms of discrimination and accuracy, while the FNNs had better calibration performances.

Table~\ref{tab:Sett1P5C501} also presents the results of models trained on an expanded cohort of patients ($n=10,000$). The results remain largely consistent between $n=1,000$ and $n=10,000$. For the machine learning methods, both the $\text{C}$-statistics and $\text{RMSE}$  show only minor improvements compared to models trained on a smaller cohort. Meanwhile, $\text{C}_{\text{benefit}}$ decreases slightly, whereas $\text{E50}_{\text{benefit}}$ and $\text{ICI}_{\text{benefit}}$  exhibit slight increases.

In Table~\ref{tab:Sett1P5C502}, $\mathrm{RMSE}^{\mathrm{SP}}$ and $\mathrm{RMSE}^{\mathrm{RMST}}$ values were generally small across all models, with ALASSO achieving the lowest $\mathrm{RMSE}^{\mathrm{SP}}$ and CSF the lowest $\mathrm{RMSE}^{\mathrm{RMST}}$, suggesting reasonable agreement between model predictions and the oracle on both scales. RMST-scale errors were uniformly smaller and less variable than SP-scale errors, consistent with the restricted mean integrating over $[0,t^*]$ and smoothing the point-wise variability of $\hat{\theta}^{\mathrm{SP}}(t)$. All $\mathrm{AUTOC}^{\mathrm{SP}}$ and $\mathrm{AUTOC}^{\mathrm{RMST}}$ values were close to zero and slightly negative, indicating that no model ranked patients by predicted benefit better than random. Under this scenario, almost all patients received a strictly positive predicted benefit, meaning the predicted-harm groups ($G_3$ and $G_4$) were nearly empty across iterations, and PB metrics were therefore not computed.

We also compared the models at different horizon points (Supplemental Material Table~\ref{tab:Sett1P75C50}) and for different levels of censoring (Supplemental Material Table~\ref{tab:Sett1P5C20}, Setting 1). All the results were improved when the censoring rate was lower (20\% instead of 50\%). Regarding the choice of the horizon time, the oracle values decreased over time, while the models performed similarly between the first and third quartiles of the time range.

The models were also compared using a matching procedure based on patients’ characteristics (Supplemental Material Table~\ref{tab:Sett1P5C20}, Setting 2). All the models performed worse compared to using a matching procedure based on predicted treatment benefit. As suggested by \cite{Klaveren2017}, this is expected as the predicted treatment benefit for $2$ patients within a pair matched by patient characteristics tends to be less similar.

%\begin{sidewaystable}
\begin{center}
\begin{table*}[ht]%
\caption{Average value at median time across the 100 simulation sets for data generation process 1.}\label{tab:Sett1P5C501}
\resizebox{\linewidth}{!}{
\scriptsize
\begin{tabular*}{\textwidth}{@{\extracolsep\fill}lcccccccc@{}}
\toprule
&\multicolumn{4}{@{}c}{\textbf{$n=1,000$}} & \multicolumn{4}{@{}c}{\textbf{$n=10,000$}} \\\cmidrule{2-5}\cmidrule{6-9}
\textbf{Model} & \textbf{C}  & \textbf{$\text{C}_{\text{benefit}}$}  & \textbf{$\text{E50}_{\text{benefit}}$}  & \textbf{$\text{ICI}_{\text{benefit}}$} &\textbf{C}  & \textbf{$\text{C}_{\text{benefit}}$}  & \textbf{$\text{E50}_{\text{benefit}}$}  & \textbf{$\text{ICI}_{\text{benefit}}$}  \\
\midrule
Oracle & $\underset{(\pm0.013)}{0.613}$ & $\underset{(\pm0.025)}{0.562}$  &    &&  $\underset{(\pm0.004)}{0.615}$  & $\underset{(\pm0.009)}{0.557}$ &   &   \\
ALASSO & $\underset{(\pm0.014)}{0.567}$  & $\underset{(\pm0.028)}{0.504}$ & $\underset{(\pm0.062)}{0.373}$ & $\underset{(\pm0.059)}{0.373}$    & $\underset{(\pm0.004)}{0.568}$  & $\underset{(\pm0.008)}{0.499}$  & $\underset{(\pm0.021)}{0.374}$ & $\underset{(\pm0.021)}{0.376}$  \\
CoxCC & $\underset{\pm0.018)}{0.545}$   &$\underset{(\pm0.043)}{0.546}$ &$\underset{(\pm0.117)}{0.321}$ &$\underset{(\pm0.116)}{0.321}$ & $\underset{(\pm0.008)}{0.581}$   &$\underset{(\pm0.021)}{0.543}$ & $\underset{(\pm0.0042)}{0.425}$ &  $\underset{(\pm0.042)}{0.42}$\\
CoxTime &  $\underset{(\pm0.016)}{0.543}$ &  $\underset{(\pm0.04)}{0.541}$&  $\underset{(\pm0.107)}{0.320}$&$\underset{(\pm0.015)}{0.325}$ &  $\underset{(\pm0.006)}{0.581}$ &$\underset{(\pm0.02)}{0.538}$ & $\underset{(\pm0.036)}{0.435}$  &$\underset{(\pm0.037)}{0.431}$  \\
IF &   $\underset{(\pm0.015)}{0.575}$  &$\underset{(\pm0.04)}{\mathbf{0.561}}$ &  $\underset{(\pm0.036)}{0.468}$ &$\underset{(\pm0.035)}{0.452}$  & $\mathbf{\underset{(\pm0.004)}{0.601}}$  & $\underset{(\pm0.011)}{0.539}$ & $\underset{(\pm0.015)}{0.473}$ &$\underset{(\pm0.013)}{0.479}$\\
CSF & $\underset{(\pm0.014)}{\mathbf{0.578}}$ & $\underset{(\pm0.023)}{0.511}$ & $\underset{(\pm0.032)}{\mathbf{0.067}}$ & $\underset{(\pm0.029)}{\mathbf{0.074}}$ & $\underset{(\pm0.004)}{0.597}$ & $\underset{(\pm0.007)}{\mathbf{0.544}}$ & $\underset{(\pm0.016)}{\mathbf{0.044}}$ & $\underset{(\pm0.013)}{\mathbf{0.044}}$ \\
\bottomrule
\end{tabular*}}
\begin{tablenotes}
\footnotesize
\item The value in brackets is the standard deviation across the 100 simulation sets. The highest C values and lowest $\text{E}_{\text{benefit}}$, $\text{ICI}_{\text{benefit}}$ and RMSE values are in bold. Patients are matched on predicted benefit. $50\%$ censoring.
\end{tablenotes}
\end{table*}
\end{center}
%\end{sidewaystable}

%\begin{sidewaystable}
\begin{center}
\begin{table*}[ht]%
\caption{Average value at median time across the 100 simulation sets for data generation process 1.}\label{tab:Sett1P5C502}
\resizebox{\linewidth}{!}{
\scriptsize
\begin{tabular*}{\textwidth}{@{\extracolsep\fill}lcccc@{}}
\toprule
&\multicolumn{4}{@{}c}{\textbf{$n=1,000$}} \\\cmidrule{2-5}
\textbf{Model} & \textbf{$\text{RMSE}^{\text{SP}}$}  & \textbf{$\text{RMSE}^{\text{RMST}}$}  & \textbf{$\text{AUTOC}^{\text{SP}}$}  & \textbf{$\text{AUTOC}^{\text{RMST}}$}  \\
\midrule
ALASSO  & $\underset{(\pm0.026)}{\mathbf{0.118}}$ & $\underset{(\pm0.009)}{0.074}$  & $\underset{(\pm0.047)}{-0.013}$   &  $\underset{(\pm0.047)}{-0.015}$ \\
CoxCC  & $\underset{(\pm0.041)}{0.212}$ & $\underset{(\pm0.013)}{0.093}$ &  $\underset{(\pm0.047)}{-0.003}$  &  $\underset{(\pm0.048)}{-0.003}$ \\
CoxTime & $\underset{(\pm0.045)}{0.217}$ & $\underset{(\pm0.011)}{0.094}$ & $\underset{(\pm0.046)}{\mathbf{-0.001}}$   &  $\underset{(\pm0.047)}{\mathbf{-0.002}}$ \\
IF & $\underset{(\pm0.025)}{0.195}$ & $\underset{(\pm0.007)}{0.085}$ &  $\underset{(\pm0.046)}{-0.017}$  &  $\underset{(\pm0.048)}{-0.016}$ \\
CSF & $\underset{(\pm0.026)}{0.135}$ & $\underset{(\pm0.008)}{\mathbf{0.072}}$ &  $\underset{(\pm0.046)}{-0.018}$  & $\underset{(\pm0.054)}{-0.020}$  \\
\bottomrule
\end{tabular*}}
\begin{tablenotes}
\footnotesize
\item The value in brackets is the standard deviation across the 100 simulation sets. The lowest RMSE values and highest AUTOC values are in bold. Patients are matched on predicted benefit. $50\%$ censoring.
\end{tablenotes}
\end{table*}
\end{center}
%\end{sidewaystable}

 \subsubsection{Data generation process 2}\label{subsubsec413}

\begin{figure*}[!ht]
\centering
\includegraphics[width=0.6\textwidth]{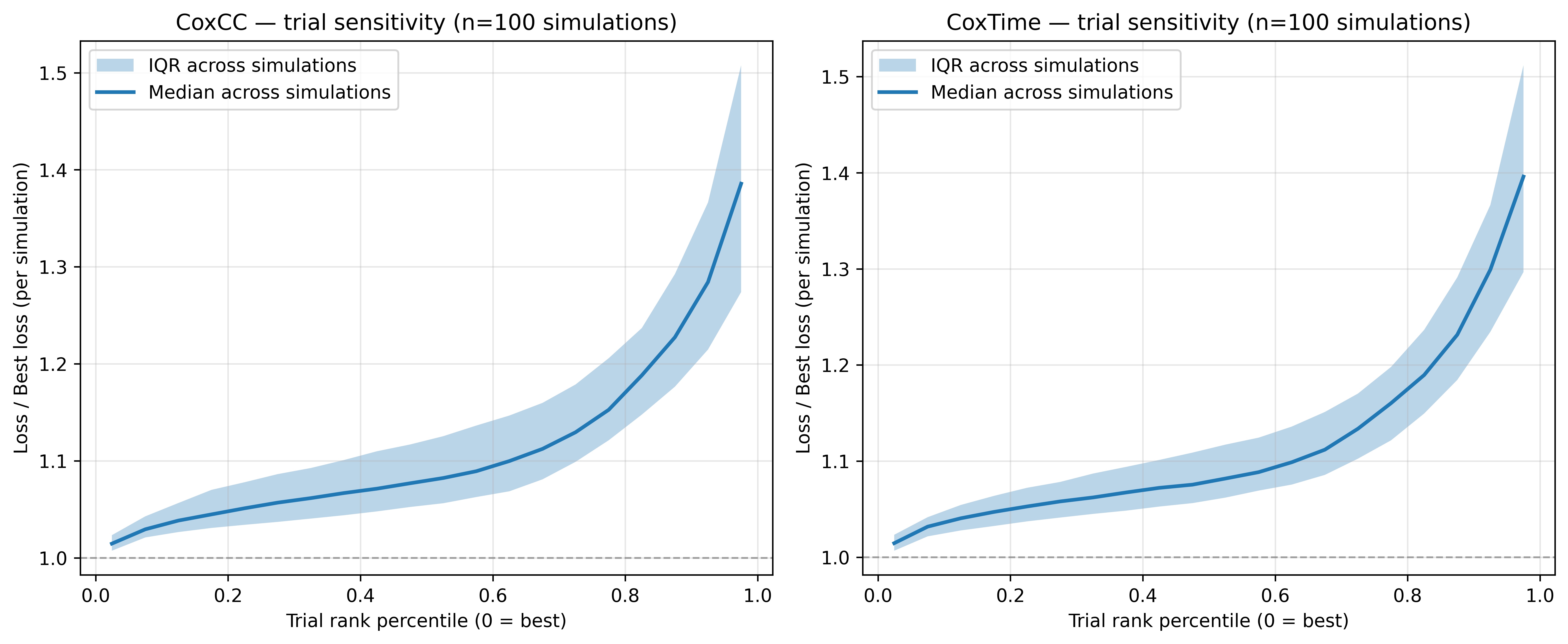}
\caption{Trial-level sensitivity of the validation loss for CoxCC (left) and CoxTime (right) for simulation setting 1 and $n=1,000$.}\label{fig:Sensibilty}
\end{figure*}

For the simulation study using the Friedman data generation process, we first describe hyperparameter search observed across the 100 simulations. For ALASSO, the penalized candidate space comprised 10 terms: the 5 main biomarker effects and their 5 biomarker-by-treatment interactions. Across the 100 simulations, the median number of penalized variables retained at the optimal $\lambda$ was 7 (Interquartile range [IQR]: 5.75--8; range: 2--9), i.e.\ 70\% of the candidate space on average. All 10 distinct candidate terms were selected in at least one simulation, corresponding to full exploration of the candidate space across replicates. Of these, 3 were selected in at least 90\%, and 7 in at least 50\%, indicating that ALASSO consistently recovers a smaller core of strongly associated terms in the great majority of replicates. Hyperparameter selection across the 100 simulation sets converged on similar architectures for CoxCC and CoxTime: a median of 3 fully-connected layers with approximately 95 neurons each, ReLU activation (selected in 91\% and 98\% of simulations respectively), moderate $L_2$ weight decay (median $5\text{--}9 \times 10^{-3}$), and low dropout (median 0.05--0.08). CoxTime preferred slightly smaller learning rates and stronger dropout than CoxCC, consistent with the higher expressivity of its time-dependent formulation. SGD with cosine annealing was the most-selected optimizer for CoxCC (56\%), whereas CoxTime more often used adaptive optimizers (Adam and Adam-AMSGrad combined: 52\%). On Figure~\ref{fig:Sensibilty}, for each of the 200 TPE trials per simulation, the validation loss is normalized by the best trial of that simulation; trials are then ordered by rank percentile. The solid line shows the median across the 100 simulated datasets and the shaded band the interquartile range. The trial-level sensitivity analysis revealed a broad basin of near-optimal hyperparameter configurations: the top 60\% of trials achieved validation losses within 5--7\% of the best trial (median across simulations), while the worst 10\% degraded by 25--40\%. This indicates that performance was robust within the explored neighborhood of good hyperparameters.

Table~\ref{tab:Sett2P5C501} reports the performance metrics of the models across the 100 simulations. CoxCC, CoxTime, IF, and CSF exhibit comparable performances, with $\text{C}$ values close to each other. CoxTime achieves the highest $\text{C}_{\text{benefit}}$  at $0.616$, with CoxCC and IF closely following. CSF yields the lowest $\text{E50}_{\text{benefit}}$  and $\text{ICI}_{\text{benefit}}$. ALASSO consistently shows less good performance metrics compared to the other models. 

Table~\ref{tab:Sett2P5C501} also presents the results for the model trained on a larger sample size. The results demonstrate only subtle differences across the evaluated metrics compared to the results obtained with the smaller sample size.

In Table~\ref{tab:Sett2P5C502}, IF achieved the lowest average RMSE on both scales, followed by ALASSO and CSF, while the two FNNs had the largest errors. All $\text{AUTOC}$ values were close to zero and mostly slightly negative, indicating that no model reliably ranked patients by predicted benefit better than random under this scenario. Regarding the PB metrics, CoxCC and CoxTime obtained the highest $\text{PB}^{\text{SP}}_{\text{M vs }\bar{\text{M}}}$ and $\text{PB}^{\text{RMST}}_{\text{M vs }\bar{\text{M}}}$ values, though these came with relatively large standard errors, suggesting high variability across simulations. For $\text{PB}^{\text{SP}}_{\text{M vs Trt}}$, differences between models were small across both scales, with ALASSO marginally leading on the SP scale and RMST scale, and all other models returning values close to zero.

%\begin{sidewaystable}
\begin{center}
\begin{table*}[ht]%
\caption{Average value at median time across the 100 simulation sets for data generation process 2.}\label{tab:Sett2P5C501}
\resizebox{\linewidth}{!}{
\scriptsize
\begin{tabular*}{\textwidth}{@{\extracolsep\fill}lcccccccc@{}}
\toprule
&\multicolumn{4}{@{}c}{\textbf{$n=1,000$}} & \multicolumn{4}{@{}c}{\textbf{$n=10,000$}} \\\cmidrule{2-5}\cmidrule{6-9}
\textbf{Model} & \textbf{C}  & \textbf{$\text{C}_{\text{benefit}}$}  & \textbf{$\text{E50}_{\text{benefit}}$}  & \textbf{$\text{ICI}_{\text{benefit}}$} &\textbf{C}  & \textbf{$\text{C}_{\text{benefit}}$}  & \textbf{$\text{E50}_{\text{benefit}}$}  & \textbf{$\text{ICI}_{\text{benefit}}$}   \\
\midrule
Oracle &  $\underset{(\pm0.038)}{0.841}$ & $\underset{(\pm0.047)}{0.583}$&     & & $\underset{(\pm0.039)}{0.838}$ & $\underset{(\pm0.039)}{0.584}$  & & \\
ALASSO &$\underset{(\pm0.093)}{0.734}$ &$\underset{(\pm0.047)}{0.583}$  &  $\underset{(\pm0.146)}{0.328}$ &  $\underset{(\pm0.136)}{0.332}$  &$\underset{(\pm0.077)}{0.719}$  &$\underset{(\pm0.035)}{0.583}$  &$\underset{(\pm0.144)}{0.326}$  & $\underset{(\pm0.129)}{0.337}$   \\
CoxCC & $\mathbf{\underset{(\pm0.048)}{0.876}}$  &$\underset{(\pm0.041)}{0.611}$  & $\underset{(\pm0.071)}{0.143}$  &$\underset{(\pm0.058)}{0.164}$   &  $\mathbf{\underset{(\pm0.046)}{0.897}}$ &$\mathbf{\underset{(\pm0.039)}{0.599}}$ &$\underset{(\pm0.070)}{0.148}$ &$\underset{(\pm0.064)}{0.168}$ \\
CoxTime & $\underset{(\pm0.057)}{0.868}$ & $\mathbf{\underset{(\pm0.042)}{0.616}}$ & $\underset{(\pm0.061)}{0.134}$  & $\underset{(\pm0.053)}{0.153}$ & $\underset{(\pm0.048)}{0.896}$ &$\underset{(\pm0.036)}{0.598}$  &$\underset{(\pm0.068)}{0.136}$&$\underset{(\pm0.060)}{0.153}$ \\
IF  & $\underset{(\pm0.051)}{0.875}$ & $\underset{(\pm0.037)}{0.608}$  & $\underset{(\pm0.090)}{0.146}$   & $\underset{(\pm0.079)}{0.172}$  &  $\underset{(\pm0.051)}{0.856}$ & $\underset{(\pm0.042)}{0.577}$ & $\underset{(\pm0.116)}{0.159}$ &$\underset{(\pm0.111)}{0.185}$ \\
CSF & $\underset{(\pm0.055)}{0.864}$ & $\underset{(\pm0.031)}{0.54}$ & $\underset{(\pm0.018)}{\mathbf{0.044}}$ & $\underset{(\pm0.017)}{\mathbf{0.057}}$ & $\underset{(\pm0.045)}{0.883}$ & $\underset{(\pm0.024)}{0.551}$ & $\underset{(\pm0.018)}{\mathbf{0.029}}$ & $\underset{(\pm0.018)}{\mathbf{0.038}}$ \\
\bottomrule
\end{tabular*}}
\begin{tablenotes}%%[341pt]
\footnotesize
\item The value in brackets is the standard deviation across the 100 simulation sets. The highest C values and lowest $\text{E}_{\text{benefit}}$ and RMSE values are in bold. Patients are matched on predicted benefit. $50\%$ censoring.
\end{tablenotes}
\end{table*}
\end{center}
%\end{sidewaystable}

%\begin{sidewaystable}
\begin{center}
\begin{table*}[ht]%
\caption{Average value at median time across the 100 simulation sets for data generation process 2.}\label{tab:Sett2P5C502}
\resizebox{\linewidth}{!}{
\scriptsize
\begin{tabular*}{\textwidth}{@{\extracolsep\fill}lcccccccc@{}}
\toprule
&\multicolumn{8}{@{}c}{\textbf{$n=1,000$}} \\\cmidrule{2-9}
\textbf{Model} & \textbf{$\text{RMSE}^{\text{SP}}$}  & \textbf{$\text{RMSE}^{\text{RMST}}$}  & \textbf{$\text{AUTOC}^{\text{SP}}$}  & \textbf{$\text{AUTOC}^{\text{RMST}}$} & $\text{PB}^{\text{SP}}_{\text{M vs }\bar{\text{M}}}$ & $\text{PB}^{\text{RMST}}_{\text{M vs }\bar{\text{M}}}$ & $\text{PB}^{\text{SP}}_{\text{M vs Trt}}$ & $\text{PB}^{\text{RMST}}_{\text{M vs Trt}}$ \\
\midrule
ALASSO  & $\underset{(\pm 0.109)}{0.163}$ & $\underset{(\pm 1.653)}{1.554}$ & $\underset{(\pm 0.041)}{\mathbf{-0.005}}$ & $\underset{(\pm 0.041)}{\mathbf{0.003}}$ & $\underset{(\pm 0.027)}{0.022}$ & $\underset{(\pm 0.064)}{0.057}$ & $\underset{(\pm 0.013)}{\mathbf{0.005}}$ & $\underset{(\pm 0.054)}{\mathbf{0.021}}$ \\
CoxCC   & $\underset{(\pm 0.113)}{0.209}$ & $\underset{(\pm 1.805)}{1.878}$ & $\underset{(\pm 0.041)}{-0.012}$ & $\underset{(\pm 0.039)}{-0.004}$ & $\underset{(\pm 0.0057)}{\mathbf{0.049}}$ & $\underset{(\pm 0.184)}{\mathbf{0.134}}$ & $\underset{(\pm 0.011)}{0.000}$ & $\underset{(\pm 0.034)}{0.003}$ \\
CoxTime & $\underset{(\pm 0.120)}{0.208}$ & $\underset{(\pm 1.856)}{1.847}$ & $\underset{(\pm 0.041)}{-0.020}$ & $\underset{(\pm 0.043)}{-0.008}$ & $\underset{(\pm 0.052)}{0.045}$ & $\underset{(\pm 0.178)}{0.122}$ & $\underset{(\pm 0.013)}{0.000}$ & $\underset{(\pm 0.036)}{0.001}$ \\
IF      & $\underset{(\pm 0.117)}{\mathbf{0.150}}$ & $\underset{(\pm 1.667)}{\mathbf{1.468}}$ & $\underset{(\pm 0.044)}{-0.023}$ & $\underset{(\pm 0.040)}{-0.012}$ & $\underset{(\pm 0.027)}{0.024}$ & $\underset{(\pm 0.073)}{0.061}$ & $\underset{(\pm 0.015)}{0.003}$ & $\underset{(\pm 0.051)}{0.015}$ \\
CSF     & $\underset{(\pm 0.115)}{0.174}$ & $\underset{(\pm 1.771)}{1.647}$ & $\underset{(\pm 0.045)}{-0.019}$ & $\underset{(\pm 0.044)}{-0.007}$ & $\underset{(\pm 0.028)}{0.026}$ & $\underset{(\pm 0.086)}{0.068}$ & $\underset{(\pm 0.014)}{0.003}$ & $\underset{(\pm 0.045)}{0.013}$ \\
\bottomrule
\end{tabular*}}
\begin{tablenotes}
\footnotesize
\item The value in brackets is the standard deviation across the 100 simulation sets. The lowest RMSE values and highest AUTOC and PB values are in bold. Patients are matched on predicted benefit. Patients are matched on predicted benefit. $50\%$ censoring.
\end{tablenotes}
\end{table*}
\end{center}
%\end{sidewaystable}

\subsection{Results on the three cancer data sets}\label{subsec42}
FNN-based methods, IF, CSF and ALASSO were applied to $3$ cancer clinical trial data datasets. For the first two datasets, which involved early breast cancer, all comparison measures were computed on the test set at horizons of 2 and 5 years, as these are standard time points of interest for clinical investigators. The third dataset consisted of renal carcinoma patients, with comparison measures assessed at 1 year, as the endpoint is progression-free survival (PFS) in the advanced setting. We applied bootstrap resampling $M=100$ times to each dataset and computed confidence intervals for all evaluation measures.

\subsubsection{Conditional average treatment effects of taxane chemotherapy}\label{subsubsec421}

The bootstrapped evaluation measures  at 2 and 5 years are presented in Table~\ref{tab:BreastBoot}. 

Overall, the neural network–based models adapted for survival, CoxCC and CoxTime, achieved the highest concordance values at 2-years, indicating superior discriminative ability compared to ALASSO and IF. When focusing on benefit-specific measures, IF slightly outperformed other models in terms of $\text{C}_{\text{benefit}}$, $\text{ICI}_{\text{benefit}}$, and $\text{E50}_{\text{benefit}}$, at both 2 and 5 years, suggesting it more accurately captures treatment effect heterogeneity.  These results suggest that while neural networks are effective at predicting overall survival in this case study, they may not fully capture treatment interactions. In contrast, IF, designed to model individualized treatment effects, better identifies who benefits most from taxane chemotherapy.

\begin{center}
\begin{table*}[ht]%
\caption{Bootstrapped evaluation measures at 2 and 5 years for the meta-analysis in the effect of taxane chemotherapy. - matching on predicted benefit.}\label{tab:BreastBoot}
\resizebox{\linewidth}{!}{
\scriptsize
\begin{tabular*}{\textwidth}{@{\extracolsep\fill}lcccccccc@{}}
\toprule
&\multicolumn{4}{@{}c}{\textbf{$t=2$}} & \multicolumn{4}{@{}c}{\textbf{$t=5$}} \\\cmidrule{2-5}\cmidrule{6-9}
\textbf{Model} & \textbf{C}  & \textbf{$\text{C}_{\text{benefit}}$}  & \textbf{$\text{E50}_{\text{benefit}}$}  & \textbf{$\text{ICI}_{\text{benefit}}$} &\textbf{C}  & \textbf{$\text{C}_{\text{benefit}}$}  & \textbf{$\text{E50}_{\text{benefit}}$}  & \textbf{$\text{ICI}_{\text{benefit}}$}   \\
\midrule
ALASSO & $\underset{[0.499, 0.563]}{0.511}$ & $\underset{[0.514, 0.750]}{0.650}$ & $\underset{[0.038, 0.212]}{0.127}$ & $\underset{[0.063, 0.206]}{0.138}$ &  $\underset{[0.15, 0.905]}{0.563}$  & $\underset{[0.491, 0.758]}{0.632}$ & $\underset{[0.032, 0.19]}{\mathbf{0.099}}$ & $\underset{[0.062, 0.189]}{0.117}$\\
CoxCC  & $\underset{[0.632, 0.714]}{\mathbf{0.673}}$ & $\underset{[0.515, 0.743]}{0.633}$  & $\underset{[0.047, 0.29]}{0.164}$  & $\underset{[0.089, 0.265]}{0.173}$  & $\underset{[0.214, 0.809]}{0.552}$ & $\underset{[0.456, 0.711]}{0.591}$ & $\underset{[0.04, 0.23]}{0.119}$& $\underset{[0.068, 0.21]}{0.132}$  \\
CoxTime& $\underset{[0.629, 0.713]}{\mathbf{0.673}}$  & $\underset{[0.482, 0.753]}{0.616}$   & $\underset{[0.059, 0.297]}{0.168}$  & $\underset{[0.09, 0.273]}{0.174}$  & $\underset{[0.25, 0.787]}{\mathbf{0.565}}$ & $\underset{[0.460, 0.692]}{0.577}$ & $\underset{[0.045, 0.23]}{0.123}$ & $\underset{[0.062, 0.218]}{0.131}$  \\
IF & $\underset{[0.454, 0.577]}{0.506}$ & $\underset{[0.554, 0.760]}{\mathbf{0.656}}$  & $\underset{[0.029, 0.203]}{0.103}$ & $\underset{[0.047, 0.19]}{0.109}$  & $\underset{[0.127, 0.891]}{0.505}$& $\underset{[0.551, 0.761]}{\mathbf{0.652}}$ & $\underset{[0.033, 0.174]}{\mathbf{0.099}}$& $\underset{[0.056, 0.181]}{\mathbf{0.106}}$  \\
CSF & $\underset{[0.457, 0.569]}{0.508}$ &$\underset{([0.405, 0.603]}{0.497}$ &$\underset{([0.024, 0.158]}{\mathbf{0.078}}$ & $\underset{[0.044, 0.166]}{\mathbf{0.093}}$ & $\underset{[0.177, 0.882]}{0.520}$ & $\underset{[0.403, 0.577]}{0.495}$ & $\underset{[0.034, 0.201]}{0.102}$ &$\underset{[0.049, 0.198]}{0.116}$\\
\bottomrule
\end{tabular*}}
\begin{tablenotes}
\footnotesize
\item  The highest C values and lowest $\text{E}_{\text{benefit}}$, $\text{ICI}_{\text{benefit}}$ and RMSE values are in bold.
\end{tablenotes}
\end{table*}
\end{center}

 Table~\ref{tab:BreastBenef} presents the predictive accuracy and treatment-prioritization performance of the five models at 2 years, using both survival-probability and restricted mean survival time estimates. Among the methods, IF showed a strong positive ranking ability whose confidence interval clearly excluded zero on the survival scale, with other models had a ranking ability close to zero. The predicted-benefit contrasts were small, and their confidence intervals included zero for every model. This suggests limited but detectable treatment-effect heterogeneity captured only by IF. For the Population Benefit metrics, CoxTime consistently yielded the highest values across all comparisons, though their confidence intervals also included zero. Overall, these results suggest limited treatment-effect heterogeneity in this RCT, with only IF capturing a detectable signal.

 \begin{center}
\begin{table*}[ht]%
\caption{Bootstrapped evaluation measures at 2 years for the meta-analysis in the effect of taxane chemotherapy. - matching on predicted benefit.}\label{tab:BreastBenef}
\resizebox{\linewidth}{!}{
\scriptsize
\begin{tabular*}{\textwidth}{@{\extracolsep\fill}lcccccc@{}}
\toprule
\textbf{Model} & \textbf{$\text{AUTOC}^{\text{SP}}$}  &  \textbf{$\text{AUTOC}^{\text{RMST}}$}  &$\text{PB}^{\text{SP}}_{\text{M vs }\bar{\text{M}}}$ & $\text{PB}^{\text{RMST}}_{\text{M vs }\bar{\text{M}}}$ & $\text{PB}^{\text{SP}}_{\text{M vs Trt}}$ & $\text{PB}^{\text{RMST}}_{\text{M vs Trt}}$  \\
\midrule
ALASSO  & $\underset{[-0.247,0.114]}{-0.052}$ & $\underset{[-0.249,0.109]}{-0.054}$ & $\underset{[-0.061,0.059]}{-0.001}$ & $\underset{[-0.069,0.065]}{-0.004}$ & $\underset{[-0.072,0.030]}{-0.015}$ & $\underset{[-0.084,0.035]}{-0.016}$ \\
CoxCC   & $\underset{[-0.101,0.093]}{0.001}$ & $\underset{[-0.097,0.095]}{0.003}$ & $\underset{[-0.042,0.095]}{0.033}$ & $\underset{[-0.030,0.106]}{0.034}$ & $\underset{[-0.055,0.033]}{-0.007}$ & $\underset{[-0.059,0.031]}{-0.010}$ \\
CoxTime & $\underset{[-0.099,0.112]}{0.001}$ & $\underset{[-0.103,0.108]}{0.001}$ & $\underset{[-0.030,0.093]}{\mathbf{0.035}}$ & $\underset{[-0.025,0.103]}{\mathbf{0.035}}$ & $\underset{[-0.036,0.027]}{\mathbf{-0.006}}$ & $\underset{[-0.040,0.026]}{\mathbf{-0.008}}$ \\
IF      & $\underset{[0.026,0.230]}{\mathbf{0.155}}$ & $\underset{[-0.014,0.226]}{\mathbf{0.136}}$ & $\underset{[-0.056,0.070]}{0.004}$ & $\underset{[-0.069,0.080]}{0.000}$ & $\underset{[-0.029,0.016]}{-0.008}$ & $\underset{[-0.034,0.018]}{-0.010}$ \\
CSF     & $\underset{[-0.084,0.120]}{0.017}$ & $\underset{[-0.087,0.115]}{0.017}$ & $\underset{[-0.056,0.050]}{0.003}$ & $\underset{[-0.071,0.063]}{0.003}$ & $\underset{[-0.043,0.017]}{-0.012}$ & $\underset{[-0.050,0.024]}{-0.012}$ \\
\bottomrule
\end{tabular*}}
\begin{tablenotes}
\footnotesize
\item  The highest values are in bold.
\end{tablenotes}
\end{table*}
\end{center}
 
 With regards to variable importance, no single variable appeared consistently in the top 5 across all model families. The neural-network models showed limited agreement with each other and with the penalized and tree-based approaches, likely reflecting the high dimensionality of the genomic feature space relative to sample size. The absence of cross-model consensus, combined with the generally low $P_{\text{top5}}$  values and wide rank confidence intervals, suggests that the treatment-effect modification signal in this dataset is weak or diffuse across the genomic feature space (Supplemental Material Table~\ref{tab:table18}).

\subsubsection{Conditional average treatment effects of adjuvant trastuzumab}\label{subsubsec422}
For the second breast cancer data set, we observed in Table~\ref{tab:TrastuBoot} that CoxCC and CoxTime reached the highest concordance values. Regarding treatment benefit, the machine-learning-based methods obtained higher values than ALASSO. CoxCC and CoxTime achieved the lowest $\text{E50}_{\text{benefit}}$ and $\text{ICI}_{\text{benefit}}$, indicating superior calibration of predicted benefit compared to ALASSO and IF. On the other hand, IF provided the highest $\text{C}_{\text{benefit}}$, suggesting it captures treatment heterogeneity more effectively, although at the cost of reduced calibration. 

In Table~\ref{tab:TrastuBenef}, IF was the only model with strong ranking ability, with AUTOC confidence intervals excluding zero on both, indicating a statistically significant ability to prioritize patients who benefit most from treatment. All other models returned AUTOC values close to zero with intervals straddling zero, suggesting no reliable treatment-effect ranking. For $\text{PB}_{M \text{ vs } \bar{M}}$, CoxCC, CoxTime, and CSF produced confidence intervals excluding zero, indicating that following the model's treatment recommendations (treating patients with positive predicted benefit and withholding treatment from those with negative predicted benefit) led to better observed survival outcomes than going against them. Notably, CSF achieved the highest and most precise $\text{PB}_{M \text{ vs } \bar{M}}^{\text{SP}}$, suggesting a sharper separation between prioritized and non-prioritized patients. In contrast, $\text{PB}_{M \text{ vs } \text{Trt}}$ remained close to zero for most models, with LASSO and IF showing negative values with intervals excluding zero, suggesting these models withhold treatment from patients who would have benefited from it. Overall, IF and CSF provide the most consistent evidence of detectable treatment-effect heterogeneity in this cohort, with IF excelling at ranking patients by benefit and CSF at identifying a well-separated high-benefit subgroup.

With regards to variable importance, cross-model consensus was limited, though a small number of variables appeared in the top 5 under more than one model family. Overall, $P_{\text{top5}}$ values remained modest across all models, and no variable achieved consistent top-5 ranking across structurally different model families. This pattern suggests that treatment-effect modification is not strongly concentrated in a small number of genomic features, and that the signal, if present, is diffuse across the high-dimensional feature space (Supplemental Material Table~\ref{tab:table19}).

\begin{center}
\begin{table*}[ht]%
\caption{Bootstrapped evaluation measures at 2 and 5 years in the RCT on adjuvant trastuzumab. - Matching on predicted benefit.}\label{tab:TrastuBoot}
\resizebox{\linewidth}{!}{
\scriptsize
\begin{tabular*}{\textwidth}{@{\extracolsep\fill}lcccccccc@{}}
\toprule
&\multicolumn{4}{@{}c}{\textbf{$t=2$}} & \multicolumn{4}{@{}c}{\textbf{$t=5$}} \\\cmidrule{2-5}\cmidrule{6-9}
\textbf{Model} & \textbf{C}  & \textbf{$\text{C}_{\text{benefit}}$}  & \textbf{$\text{E50}_{\text{benefit}}$}  & \textbf{$\text{ICI}_{\text{benefit}}$} &\textbf{C}  & \textbf{$\text{C}_{\text{benefit}}$}  & \textbf{$\text{E50}_{\text{benefit}}$}  & \textbf{$\text{ICI}_{\text{benefit}}$}   \\
\midrule
ALASSO &$\underset{[0.462, 0.549]}{0.505}$ & $\underset{[0.698, 0.756]}{0.729}$& $\underset{[0.062, 0.146]}{0.104}$& $\underset{[0.091, 0.151]}{0.12}$ & $\underset{[0.477, 0.533]}{0.503}$   & $\underset{[0.693, 0.758]}{0.729}$  &  $\underset{[0.05, 0.105]}{0.074}$ & $\underset{[0.064, 0.111]}{0.086}$\\
CoxCC  & $\underset{[0.55, 0.625]}{\mathbf{0.588}}$ & $\underset{[0.463, 0.556]}{0.512}$  & $\underset{[0.029, 0.106]}{\mathbf{0.065}}$ & $\underset{[0.036, 0.097]}{0.069}$  & $\underset{[0.546, 0.607]}{\mathbf{0.581}}$ & $\underset{[0.460, 0.563]}{0.517}$ & $\underset{[0.027, 0.084]}{\mathbf{0.058}}$ & $\underset{[0.039, 0.086]}{0.063}$ \\
CoxTime& $\underset{[0.542, 0.62]}{0.586}$  & $\underset{[0.457, 0.556]}{0.504}$  & $\underset{[0.033, 0.096]}{\mathbf{0.065}}$ & $\underset{[0.037, 0.094]}{\mathbf{0.064}}$  & $\underset{[0.555, 0.607]}{0.580}$ & $\underset{[0.473, 0.564]}{0.512}$ & $\underset{[0.021, 0.087]}{0.059}$ &$\underset{[0.037, 0.087]}{\mathbf{0.062}}$  \\
IF & $\underset{[0.468, 0.535]}{0.503}$ & $\underset{[0.715, 0.774]}{\mathbf{0.744}}$  & $\underset{[0.05, 0.141]}{0.088}$  & $\underset{[0.08, 0.14]}{0.104}$  & $\underset{[0.477, 0.534]}{0.503}$ &$\underset{[0.737, 0.787]}{\mathbf{0.762}}$ & $\underset{[0.04, 0.091]}{0.069}$ & $\underset{[0.052, 0.099]}{0.076}$  \\
CSF & $\underset{[0.463, 0.552]}{0.503}$ & $\underset{[0.468, 0.538]}{0.503}$& $\underset{[0.040, 0.116]}{0.081}$ & $\underset{[0.055, 0.113]}{0.084}$ & $\underset{[0.480, 0.527]}{0.501}$ & $\underset{[0.469, 0.535]}{0.502}$ & $\underset{[0.016, 0.081]}{0.043}$ & $\underset{[0.021, 0.085]}{0.049}$\\
\bottomrule
\end{tabular*}}
\begin{tablenotes}
\footnotesize
\item  The highest C values and lowest $\text{E}_{\text{benefit}}$, $\text{ICI}_{\text{benefit}}$ and RMSE values are in bold.
\end{tablenotes}
\end{table*}
\end{center}

 \begin{center}
\begin{table*}[ht]%
\caption{Bootstrapped evaluation measures at 2 and 5 years in the RCT on adjuvant trastuzumab. - Matching on predicted benefit.}\label{tab:TrastuBenef}
\resizebox{\linewidth}{!}{
\scriptsize
\begin{tabular*}{\textwidth}{@{\extracolsep\fill}lcccccc@{}}
\toprule
\textbf{Model} & \textbf{$\text{AUTOC}^{\text{SP}}$}  &  \textbf{$\text{AUTOC}^{\text{RMST}}$}  &$\text{PB}^{\text{SP}}_{\text{M vs }\bar{\text{M}}}$ & $\text{PB}^{\text{RMST}}_{\text{M vs }\bar{\text{M}}}$ & $\text{PB}^{\text{SP}}_{\text{M vs Trt}}$ & $\text{PB}^{\text{RMST}}_{\text{M vs Trt}}$  \\
\midrule
ALASSO  & $\underset{[-0.045,\,0.061]}{0.009}$  & $\underset{[-0.045,\,0.061]}{0.008}$  & $\underset{[-0.030,\,0.029]}{0.001}$  & $\underset{[-0.027,\,0.031]}{0.002}$  & $\underset{[-0.055,\,-0.025]}{-0.040}$ & $\underset{[-0.045,\,-0.016]}{-0.031}$ \\
CoxCC   & $\underset{[-0.038,\,0.048]}{0.006}$  & $\underset{[-0.039,\,0.048]}{0.006}$  & $\underset{[-0.044,\,0.087]}{0.074}$  & $\underset{[0.033,\,0.070]}{0.057}$   & $\underset{[-0.018,\,0.002]}{-0.004}$  & $\underset{[-0.016,\,0.003]}{-0.003}$  \\
CoxTime & $\underset{[-0.043,\,0.056]}{0.009}$  & $\underset{[-0.043,\,0.056]}{0.010}$  & $\underset{[0.031,\,0.087]}{0.066}$   & $\underset{[0.023,\,0.067]}{0.050}$   & $\underset{[-0.025,\,0.003]}{-0.007}$  & $\underset{[-0.020,\,0.003]}{-0.007}$  \\
IF      & $\underset{[0.019,\,0.099]}{\mathbf{0.057}}$ & $\underset{[0.034,\,0.115]}{\mathbf{0.072}}$ & $\underset{[-0.031,\,0.029]}{0.000}$ & $\underset{[-0.033,\,0.036]}{-0.000}$ & $\underset{[-0.057,\,-0.028]}{-0.041}$ & $\underset{[-0.047,\,-0.014]}{-0.031}$ \\
CSF     & $\underset{[-0.054,\,0.046]}{0.009}$  & $\underset{[-0.047,\,0.047]}{-0.004}$ & $\underset{[0.075,\,0.082]}{\mathbf{0.078}}$ & $\underset{[0.055,\,0.065]}{\mathbf{0.060}}$ & $\underset{[-0.003,\,0.000]}{\mathbf{-0.001}}$ & $\underset{[-0.003,\,0.001]}{\mathbf{-0.001}}$ \\
\bottomrule
\end{tabular*}}
\begin{tablenotes}
\footnotesize
\item  The highest values are in bold.
\end{tablenotes}
\end{table*}
\end{center}

\subsubsection{Conditional average treatment effects of avelumab}\label{subsubsec423}

For the advanced renal cell carcinoma cohort (Table~\ref{tab:AveluBoot}), CoxTime achieved the highest concordance at 1 year, indicating stronger discrimination overall, whereas IF obtained the highest $\text{C}_{\text{benefit}}$, suggesting a better ability to capture treatment heterogeneity. In terms of calibration of treatment benefit, CoxCC reached the lowest $\text{E50}_{\text{benefit}}$ and $\text{ICI}_{\text{benefit}}$, reflecting superior agreement between predicted and observed benefit. By contrast, IF showed higher calibration errors despite its strong discrimination.

In Table~\ref{tab:AveluBenef}, as in the previous two cancer examples, IF was the model with the best ranking ability and whose ranking ability excluded zero on both scales, while all other models had ranking abilities close to zero. None of the predicted-benefit contrasts excluded zero for any model, and the RMST-scale PB intervals were extremely wide, reflecting high variability in the restricted-mean benefit estimates at the one-year horizon in this cohort. IF provides the only reproducible signal of treatment-effect heterogeneity, whereas the predicted-benefit metrics are too unstable here to support firm conclusions.

With regards to the variable importance, 3 variables appeared consistently in the top 5 across model families. The McDermott angiogenesis signature was selected by LASSO ($0.92$), IF ($0.72$), and CSF ($0.46$), reaching a mean rank of $1.7$ under LASSO (Supplemental Material Table~\ref{tab:table20}). Patient age was the dominant modifier under both neural-network models (CoxCC: $0.93$, CoxTime: $0.91$, both with mean rank near 2) and was also picked up by IF ($0.46$). The cell-cell signaling pathway (\texttt{B9991003\_rc5\_Cell\_cell\_signaling}) was selected by CSF ($0.62$), IF ($0.42$), and LASSO ($0.31$). The agreement of structurally different model families on these three variables, despite their different bases for ranking, provides convergent evidence that they are associated with the magnitude of avelumab benefit.

\begin{center}
\begin{table*}[ht]%
\caption{Bootstrapped evaluation measures at 1 year in the RCT on avelumab. - Matching on predicted benefit.}\label{tab:AveluBoot}
\resizebox{\linewidth}{!}{
\scriptsize
\begin{tabular*}{\textwidth}{@{\extracolsep\fill}lcccc@{}}
\toprule
&\multicolumn{4}{@{}c}{\textbf{$t=1$}} 
\\\cmidrule{2-5}
\textbf{Model} & \textbf{C}  & \textbf{$\text{C}_{\text{benefit}}$}  & \textbf{$\text{E50}_{\text{benefit}}$}  & \textbf{$\text{ICI}_{\text{benefit}}$} \\
\midrule
ALASSO & $\underset{[0.341, 0.598]}{0.484}$ & $\underset{[0.582, 0.771]}{0.692}$ & $\underset{[0.045,0.289]}{0.179}$ &   $\underset{[0.079, 0.316]}{0.183}$\\
CoxCC   &  $\underset{[0.417, 0.651]}{0.544}$ &  $\underset{[0.51, 0.61]}{0.561}$ & $\underset{[0.022, 0.14]}{0.061}$ & $\underset{[0.038, 0.133]}{0.073}$ \\
CoxTime &$\underset{[0.407, 0.664]}{\mathbf{0.549}}$ & $\underset{[0.514, 0.612]}{0.563}$ & $\underset{[0.021, 0.144]}{0.068}$ & $\underset{[0.036, 0.125]}{0.078}$ \\
IF  &$\underset{[0.39, 0.639]}{0.517}$ & $\underset{[0.695, 0.759]}{\mathbf{0.728}}$ & $\underset{[0.158, 0.368]}{0.273}$ & $\underset{[0.211, 0.335]}{0.278}$\\
CSF & $\underset{[0.409, 0.636]}{0.520}$ & $\underset{[0.439, 0.554]}{0.499}$& $\underset{[0.016, 0.108]}{\mathbf{0.051}}$ & $\underset{[0.028, 0.100]}{\mathbf{0.062}}$\\
\bottomrule
\end{tabular*}}
\begin{tablenotes}
\footnotesize
\item  The highest C values and lowest $\text{E}_{\text{benefit}}$, $\text{ICI}_{\text{benefit}}$ and RMSE values are in bold.
\end{tablenotes}
\end{table*}
\end{center}

 \begin{center}
\begin{table*}[ht]%
\caption{Bootstrapped evaluation measures at 1 year in the RCT on avelumab. - Matching on predicted benefit.}\label{tab:AveluBenef}
\resizebox{\linewidth}{!}{
\scriptsize
\begin{tabular*}{\textwidth}{@{\extracolsep\fill}lcccccc@{}}
\toprule
\textbf{Model} & \textbf{$\text{AUTOC}^{\text{SP}}$}  &  \textbf{$\text{AUTOC}^{\text{RMST}}$}  &$\text{PB}^{\text{SP}}_{\text{M vs }\bar{\text{M}}}$ & $\text{PB}^{\text{RMST}}_{\text{M vs }\bar{\text{M}}}$ & $\text{PB}^{\text{SP}}_{\text{M vs Trt}}$ & $\text{PB}^{\text{RMST}}_{\text{M vs Trt}}$  \\
\midrule
ALASSO  & $\underset{[-0.070,\,0.140]}{0.023}$  & $\underset{[-0.075,\,0.138]}{0.023}$  & $\underset{[-0.095,\,0.087]}{-0.004}$ & $\underset{[-0.568,\,0.489]}{-0.037}$ & $\underset{[-0.076,\,0.013]}{-0.029}$ & $\underset{[-0.459,\,0.078]}{-0.190}$ \\
CoxCC   & $\underset{[-0.137,\,0.084]}{-0.022}$ & $\underset{[-0.138,\,0.064]}{-0.022}$ & $\underset{[-0.067,\,0.110]}{0.020}$  & $\underset{[-0.365,\,0.587]}{0.139}$  & $\underset{[-0.061,\,0.031]}{-0.018}$ & $\underset{[-0.356,\,0.133]}{-0.103}$ \\
CoxTime & $\underset{[-0.123,\,0.074]}{-0.027}$ & $\underset{[-0.125,\,0.077]}{-0.026}$ & $\underset{[-0.043,\,0.099]}{0.024}$  & $\underset{[-0.294,\,0.564]}{\mathbf{0.142}}$  & $\underset{[-0.053,\,0.024]}{-0.015}$ & $\underset{[-0.314,\,0.119]}{\mathbf{-0.098}}$ \\
IF      & $\underset{[0.065,\,0.237]}{\mathbf{0.152}}$ & $\underset{[0.076,\,0.255]}{\mathbf{0.169}}$ & $\underset{[-0.104,\,0.095]}{-0.002}$ & $\underset{[-0.481,\,0.540]}{-0.025}$ & $\underset{[-0.081,\,0.018]}{-0.029}$ & $\underset{[-0.425,\,0.097]}{-0.187}$ \\
CSF     & $\underset{[-0.080,\,0.098]}{-0.003}$ & $\underset{[-0.094,\,0.105]}{-0.002}$ & $\underset{[-0.086,\,0.095]}{\mathbf{0.027}}$ & $\underset{[-0.488,\,0.619]}{0.114}$  & $\underset{[-0.069,\,0.025]}{\mathbf{-0.014}}$ & $\underset{[-0.403,\,0.142]}{-0.113}$ \\
\bottomrule
\end{tabular*}}
\begin{tablenotes}
\footnotesize
\item  The highest values are in bold.
\end{tablenotes}
\end{table*}
\end{center}

\clearpage

\section{Discussion}\label{sec5}
In this paper, we compared two FNN methods adapted to survival outcomes (CoxCC and CoxTime) and two random forest models (IF and CSF) for their ability to predict individualized treatment benefits in the context of time-to-event data from randomized clinical trials, using a linear CoxPH model with an adaptive LASSO penalty as a benchmark method. The comparison was conducted on 2 simulation sets: one introducing nonlinear interaction with treatment and another based on Friedman's random function generator with nonlinear and high-order interactions. Additionally, the models were evaluated on 3 case studies from oncology clinical trials.

Under data generation process 1, IF achieved the best discrimination ($C_{\text{benefit}}$), while the FNNs and CSF showed superior calibration, with CSF having the lowest $\text{E50}_{\text{benefit}}$ and $\text{ICI}_{\text{benefit}}$ of all models. CSF also achieved the lowest $\text{RMSE}^{\text{RMST}}$, while ALASSO had the lowest $\text{RMSE}^{\text{SP}}$. For data generation process 2, the performances of the machine learning methods were better than the linear ALASSO method regarding the discrimination and calibration metrics, and RMSE. Overall, CSF consistently delivered the best calibration across both processes, while IF excelled in discrimination and RMSE.

Across the three cancer clinical trial datasets, the models displayed complementary strengths in predicting treatment benefit. CoxCC and CoxTime consistently achieved the highest concordance values and the lowest $\text{E50}_{\text{benefit}}$ and $\text{ICI}_{\text{benefit}}$, indicating superior overall discrimination and more reliable calibration of predicted benefit than IF. IF, on the other hand, consistently yielded the highest $C_{\text{benefit}}$ and was the only model with AUTOC confidence intervals excluding zero across all three datasets, indicating better identification of patients with larger differential treatment effects, though sometimes at the expense of calibration. CSF showed mixed performance: it achieved the lowest $\text{E50}_{\text{benefit}}$ and $\text{ICI}_{\text{benefit}}$ in the breast cancer taxane dataset and the renal carcinoma dataset, indicating strong calibration, but performed poorly on $C_{\text{benefit}}$ and AUTOC across all datasets. In the trastuzumab dataset, CSF achieved the highest and most precise $\text{PB}_{M \text{ vs } \bar{M}}^{\text{SP}}$, suggesting it effectively identifies well-separated high-benefit subgroups. ALASSO offered moderate, consistent performance across metrics. Overall, these findings reflect a trade-off: neural network--based models provide well-calibrated benefit estimates, while IF is more sensitive to heterogeneity. ALASSO may offer a consistent alternative, even if it does not fit nonlinear effects or interactions in these types of cancer trial datasets.

In this paper, we used measures specifically adapted to treatment benefit evaluation in the context of time-to-event data. The assessment of treatment benefit prediction models has been an active area of research. \cite{Efthimiou2022} developed measures of discrimination and calibration, building upon the work of \cite{Klaveren2017}, who introduced the C-for-benefit metric specifically for evaluating treatment benefits. We adapted their approach to time-to-event data by calculating this measure using survival probabilities. While C-for-benefit provides a flexible framework for comparing individualized treatment effects, its performance can vary with the matching procedure used and, as noted in recent methodological work \citep{Efthimiou2025}, it does not constitute a proper scoring rule.  Additionally, we modified the RMSE metric to assess treatment benefit for time-to-event data by evaluating the difference between predicted and true treatment benefits, both derived from survival probabilities. Using adapted measures for treatment benefit is of crucial importance, as, for example, the standard concordance index is unable to directly capture the treatment benefit. Future work could include associating uncertainty measures with conditional average treatment effects, which could, for example, be evaluated by a bootstrap method as in \cite{Roblin2024}.

One limitation of our approach concerns the matching procedure used to estimate treatment benefit. Matching on predicted benefit can introduce variability in the estimated metrics depending on the quality of the matches and the degree of overlap between treated and control patients in terms of predicted benefit. This matching-dependent variability may partly explain the substantially elevated $\text{E50}_{\text{benefit}}$ and $\text{ICI}_{\text{benefit}}$ values observed across all models in some settings. Future work should investigate double-robust estimators and methods that do not rely on explicit matching, which may provide more stable estimates of treatment benefit, particularly in settings with limited overlap or high censoring.

\cite{Weberpals2025} emphasize the significant potential of leveraging complex causal machine learning models to refine individualized treatment effect estimation. Causal machine learning provides a promising framework for enhancing precision medicine by predicting patient-specific cancer treatment responses using rich, multimodal data. Although clinical applications of causal machine learning are still emerging, its capability to integrate high-dimensional data and provide robust causal inference shows great promise for improving personalized treatment choices. Our work is indeed centered on the RCT framework, which allows for the determination of causal relationships between an intervention and its outcome.  However, the methods used in this paper could be adapted beyond RCTs and applied to real-world data cohorts by incorporating propensity scores or other causal approaches. \cite{Cui2023} proposed such an approach using causal forests, a machine learning method used to estimate heterogeneous treatment effects in observational studies. 

\section{Conclusion}\label{sec6}
Calibration of predicted benefit varied across datasets and data-generating processes, with no single model family dominating consistently. CoxCC and CoxTime showed lower calibration errors in several settings, while CSF was competitive or superior in others. Tree-based methods (particularly IF) delivered consistently superior discrimination and patient prioritization across all evaluated datasets. IF achieved the highest $C_{\text{benefit}}$  and statistically significant AUTOC values across all three cancer datasets. CSF showed strong calibration competitive with neural networks. Our adapted metrics for survival benefits ($C_{\text{benefit}}$, AUTOC, Population Benefit) uncover performance differences that standard indices miss. These findings confirm causal machine learning's promise for precision medicine: IF consistently identified who benefited most from treatment, CSF provided well-separated high-benefit subgroups, and neural networks produced absolute benefit estimates with lower point-wise error in several settings.

\section{Acknowledgements}
   Elvire Roblin acknowledges financial support by Foundation Philanthropia Lombard-Odier. The authors also acknowledge the National Surgical Adjuvant Breast and Bowel Project (NSABP) investigators of the B-31 trial who submitted data from the original study to dbGaP (dbGaP Study Accession: phs000826.v1.p1) and the NIH data repository. The B-31 trial was supported by: National Cancer Institute, Department of Health and Human Services, Public Health Service, Grants U10-CA-12027, U10-CA-69651, U10-CA-37377, and U10-CA-69974, and by a grant from the Pennsylvania Department of Health. 
   
\section{Declaration of conflicting interests}
    The authors declare no potential conflict of interests.

\clearpage

\renewcommand{\thesection}{\Alph{section}}
\setcounter{section}{0}
\clearpage 
\section*{Supplementary Material}

\section{Hyperparameter search\label{secA1}}

\begin{table}[!ht]
\centering
\caption{Hyperparameters search using Tree-Parzen Algorithm with the python package optuna.}\label{HyperparameterTable}
\begin{tabular}{@{}ll@{}}
\toprule
 \textbf{Hyperparameter} & \textbf{Value} \\ \midrule
 Activation function & $\{\text{elu, relu, tanh}\}$ \\
 Batch size & $\{8,16,32,64,128,256\}$ \\
 Dropout rate & $[0.01,0.5]$ \\
 $L_{2}$ regularization & $[0.001,0.1]$ \\
 Learning rate & $[0.001,0.01]$ \\
 Number of hidden layers & $\{1,2,3,4\}$ \\
 Number of neurons per layer & $\lbrack 4,128\rbrack$ \\
 Optimization algorithm & $\{\text{Adam, AdamAMSGRAD, RMSProp, SGDWR}\}$ \\ \bottomrule
\end{tabular}
\end{table}

\subsection{Hyperparameters in Simulation Setting 1}
Tables~\ref{tab:table12} and~\ref{tab:table13} present the selected hyperparameters across 100 simulations, reported as median and interquartile range for continuous and integer hyperparameters, and as the most frequently selected value for categorical ones.

For ALASSO on both settings ($n=1{,}000$ and $n=10{,}000$), no biomarker was selected by ALASSO in more than 50\% of the 100 replicates. The median number of penalized terms retained at the optimal $\lambda$ was 6 (IQR: $[2.75, 10]$; range: $[0, 23]$) for $n=1{,}000$ and 5 (IQR: $[0, 8.25]$; range: $[0, 21]$) for $n=10{,}000$, corresponding to only 15\% and 12\% of the 40-term candidate space, respectively. All 40 candidate terms appeared in at least one selection across replicates.

 \begin{figure*}[ht]
\centering
\includegraphics[width=0.6\textwidth]{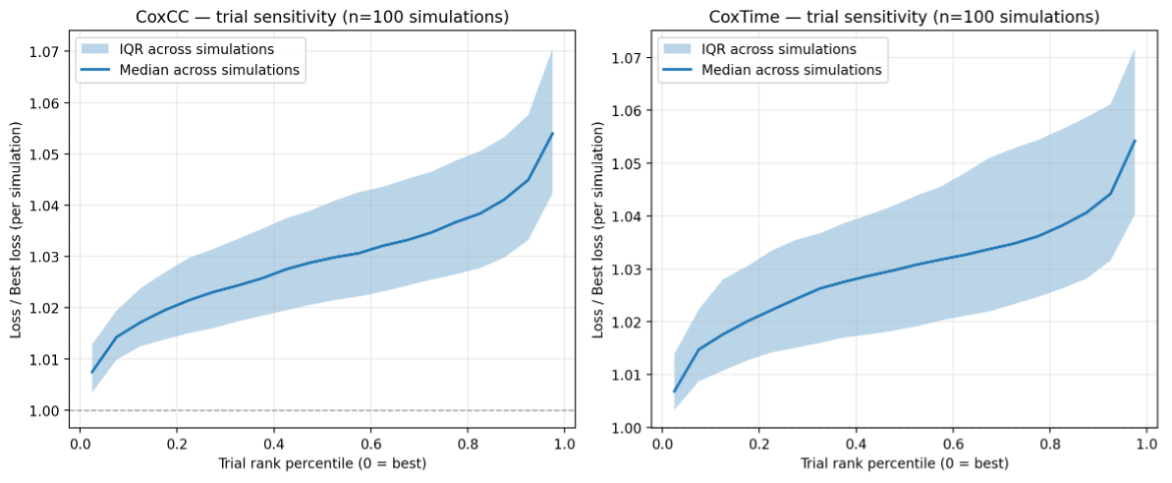}
\caption{Trial-level sensitivity of the validation loss for CoxCC (left) and CoxTime (right) for simulation setting 1 and $n=1,000$.}\label{fig:SensitivtySet11}
\end{figure*}

\clearpage

\begin{table}[htbp]
\centering
\caption{Hyperparameter selection summary across 100 simulation sets, Simulation setting 1, $n=1{,}000$.}
\label{tab:table12}
\begin{tabular}{llcc}
\toprule
\multirow{2}{*}{Model} & \multirow{2}{*}{Hyperparameter} & Continuous/Integer & Categorical \\
 & & $\underset{[\text{IQR}]}{\text{Median}}$ & Mode / Top categories \\
\midrule
\multirow{8}{*}{CoxCC}
 & activation     & ---                                                        & relu (58\%), tanh (42\%) \\
 & batch\_size    & ---                                                        & 128 (37\%), 64 (22\%), 32 (21\%) \\
 & n\_layers      & $\underset{[1,\,3]}{2}$                                     & --- \\
 & learning\_rate & $\underset{[2.64\text{e-}03,\,7.10\text{e-}03]}{4.30\text{e-}03}$ & --- \\
 & neurons        & $\underset{[42,\,98]}{76}$                                  & --- \\
 & optimizer      & ---                                                        & adam\_amsgrad (40\%), adam (29\%), rmsprop (19\%) \\
 & l2             & $\underset{[0.015,\,0.068]}{0.038}$                         & --- \\
 & dropout        & $\underset{[0.053,\,0.211]}{0.119}$                         & --- \\
\midrule
\multirow{8}{*}{CoxTime}
 & activation     & ---                                                        & relu (51\%), tanh (49\%) \\
 & batch\_size    & ---                                                        & 128 (39\%), 64 (22\%), 32 (17\%) \\
 & n\_layers      & $\underset{[1,\,3]}{2}$                                     & --- \\
 & learning\_rate & $\underset{[2.74\text{e-}03,\,6.86\text{e-}03]}{4.72\text{e-}03}$ & --- \\
 & neurons        & $\underset{[42,\,95]}{72}$                                  & --- \\
 & optimizer      & ---                                                        & adam (44\%), rmsprop (24\%), adam\_amsgrad (18\%) \\
 & l2             & $\underset{[0.014,\,0.062]}{0.036}$                         & --- \\
 & dropout        & $\underset{[0.075,\,0.221]}{0.148}$                         & --- \\
\bottomrule
\end{tabular}
\end{table}

 \begin{figure*}[ht]
\centering
\includegraphics[width=0.6\textwidth]{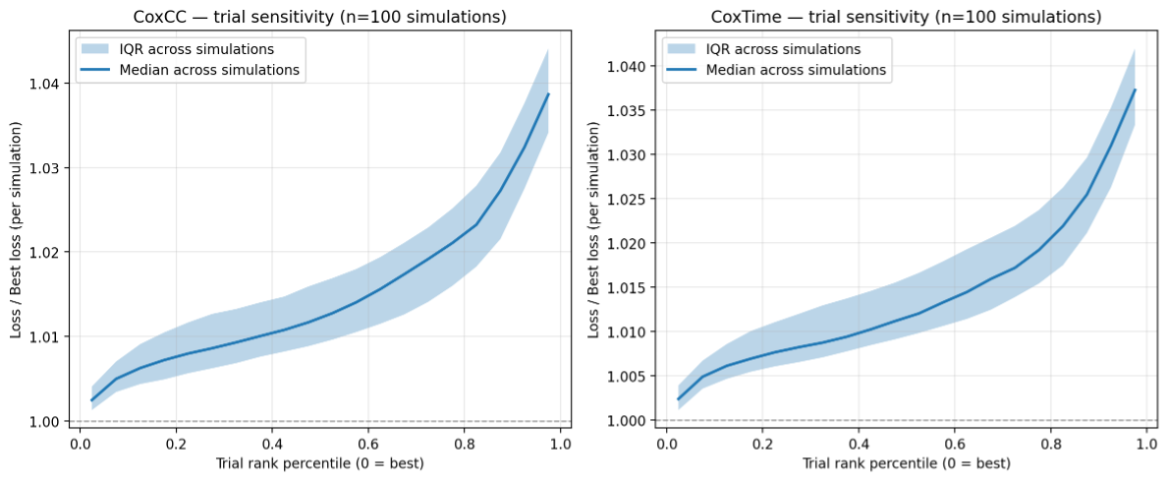}
\caption{Trial-level sensitivity of the validation loss for CoxCC (left) and CoxTime (right) for simulation setting 1 and $n=10,000$.}\label{fig:SensitivtySet12}
\end{figure*}

\begin{table}[htbp]
\centering
\caption{Hyperparameter selection summary across 100 simulation sets, Simulation setting 1, $n=10{,}000$.}
\label{tab:table13}
\begin{tabular}{llcc}
\toprule
\multirow{2}{*}{Model} & \multirow{2}{*}{Hyperparameter} & Continuous/Integer & Categorical \\
 & & $\underset{[\text{IQR}]}{\text{Median}}$ & Mode / Top categories \\
\midrule
\multirow{8}{*}{CoxCC}
 & activation     & ---                                                        & relu (99\%), tanh (1\%) \\
 & batch\_size    & ---                                                        & 128 (31\%), 64 (29\%), 32 (19\%) \\
 & n\_layers      & $\underset{[2,\,3]}{2}$                                     & --- \\
 & learning\_rate & $\underset{[2.38\text{e-}03,\,6.57\text{e-}03]}{4.91\text{e-}03}$ & --- \\
 & neurons        & $\underset{[74,\,112]}{92}$                                 & --- \\
 & optimizer      & ---                                                        & sgdwr (84\%), adam\_amsgrad (8\%), adam (5\%) \\
 & l2             & $\underset{[9.24\text{e-}03,\,0.033]}{0.019}$              & --- \\
 & dropout        & $\underset{[0.089,\,0.242]}{0.178}$                         & --- \\
\midrule
\multirow{8}{*}{CoxTime}
 & activation     & ---                                                        & relu (97\%), tanh (3\%) \\
 & batch\_size    & ---                                                        & 64 (34\%), 128 (34\%), 32 (18\%) \\
 & n\_layers      & $\underset{[1,\,3]}{2}$                                     & --- \\
 & learning\_rate & $\underset{[1.80\text{e-}03,\,6.38\text{e-}03]}{3.71\text{e-}03}$ & --- \\
 & neurons        & $\underset{[68,\,109]}{96}$                                 & --- \\
 & optimizer      & ---                                                        & sgdwr (70\%), adam (15\%), rmsprop (10\%) \\
 & l2             & $\underset{[6.56\text{e-}03,\,0.033]}{0.015}$              & --- \\
 & dropout        & $\underset{[0.093,\,0.249]}{0.165}$                         & --- \\
\bottomrule
\end{tabular}
\end{table}

\clearpage

\subsection{Hyperparameters in Simulation Setting 2}
Table~\ref{tab:table14} presents the selected hyperparameters across 100 simulations, reported as median and interquartile range for continuous and integer hyperparameters, and as the most frequently selected value for categorical ones.

We next present the results for $n=10{,}000$. For ALASSO, the penalized candidate space comprised 10 terms: the 5 main biomarker effects and their 5 biomarker-by-treatment interactions (the treatment indicator is forced unpenalized and is excluded from these counts). Across the 100 simulations, the median number of penalized variables retained at the optimal $\lambda$ was 7 (IQR: $[5, 8]$; range: $[3, 9]$), i.e.\ 70\% of the candidate space on average. A total of 10 distinct variables were selected at least once, corresponding to full exploration of the candidate space across replicates. Of these, 0 were selected in every simulation, 4 in at least 90\%, and 5 in at least 50\%.

\begin{figure*}[t]
\centering
\includegraphics[width=0.6\textwidth]{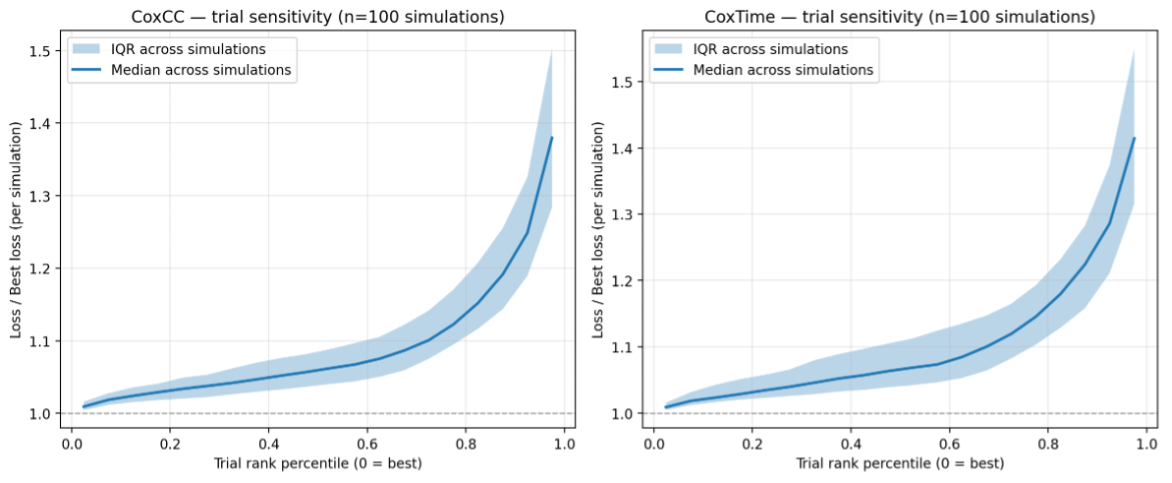}
\caption{Trial-level sensitivity of the validation loss for CoxCC (left) and CoxTime (right) for simulation setting 2 and $n=10,000$.}\label{fig:SensitivtySet22}
\end{figure*}

In Figure~\ref{fig:SensitivtySet22}, for each of the 200 TPE trials per simulation, the validation loss is normalized by the best trial of that simulation; trials are then ordered by rank percentile. The solid line shows the median across the 100 simulated datasets and the shaded band the interquartile range. The trial-level sensitivity analysis revealed a broad basin of near-optimal hyperparameter configurations for both models: the top 60\% of trials achieved validation losses within 5--8\% of the best trial (median across simulations), while the worst 10\% degraded by approximately 35--50\%. The IQR band remains narrow throughout the middle range of the distribution, indicating consistent behavior across simulations, before widening at the upper tail, reflecting greater variability among the poorest-performing trials. Overall, performance was robust within the explored neighborhood of good hyperparameter configurations, with results closely mirroring those observed at $n=1{,}000$.

\begin{table}[htbp]
\centering
\caption{Hyperparameter selection summary across 100 simulation sets under data-generating process 2.}
\label{tab:table14}
\begin{tabular}{llcc}
\toprule
\multirow{2}{*}{Model} & \multirow{2}{*}{Hyperparameter} & Continuous/Integer & Categorical \\
 & & $\underset{[\text{IQR}]}{\text{Median}}$ & Mode / Top categories \\
\midrule
\multirow{8}{*}{CoxCC}
 & activation     & ---                                                        & relu (91\%), tanh (9\%) \\
 & batch\_size    & ---                                                        & 32 (40\%), 64 (34\%), 128 (15\%) \\
 & n\_layers      & $\underset{[2,\,4]}{3}$                                     & --- \\
 & learning\_rate & $\underset{[1.96\text{e-}03,\,7.61\text{e-}03]}{4.22\text{e-}03}$ & --- \\
 & neurons        & $\underset{[71,\,113]}{95}$                                 & --- \\
 & optimizer      & ---                                                        & sgdwr (56\%), adam (20\%), adam\_amsgrad (16\%) \\
 & l2             & $\underset{[2.16\text{e-}03,\,0.018]}{8.57\text{e-}03}$    & --- \\
 & dropout        & $\underset{[0.025,\,0.097]}{0.047}$                         & --- \\
\midrule
\multirow{8}{*}{CoxTime}
 & activation     & ---                                                        & relu (98\%), tanh (2\%) \\
 & batch\_size    & ---                                                        & 64 (38\%), 32 (30\%), 128 (22\%) \\
 & n\_layers      & $\underset{[2,\,4]}{3}$                                     & --- \\
 & learning\_rate & $\underset{[1.58\text{e-}03,\,6.34\text{e-}03]}{3.24\text{e-}03}$ & --- \\
 & neurons        & $\underset{[73,\,114]}{96}$                                 & --- \\
 & optimizer      & ---                                                        & sgdwr (40\%), adam (29\%), adam\_amsgrad (23\%) \\
 & l2             & $\underset{[1.87\text{e-}03,\,0.016]}{5.10\text{e-}03}$    & --- \\
 & dropout        & $\underset{[0.037,\,0.131]}{0.080}$                         & --- \\
\bottomrule
\end{tabular}
\end{table}

\begin{table}[htbp]
\centering
\caption{Hyperparameter selection summary across 100 simulation sets under data-generating process 2, $n=10{,}000$.}
\label{tab:table15}
\begin{tabular}{llcc}
\toprule
\multirow{2}{*}{Model} & \multirow{2}{*}{Hyperparameter} & Continuous/Integer & Categorical \\
 & & $\underset{[\text{IQR}]}{\text{Median}}$ & Mode / Top categories \\
\midrule
\multirow{8}{*}{CoxCC}
 & activation     & ---                                                        & relu (100\%) \\
 & batch\_size    & ---                                                        & 128 (47\%), 64 (28\%), 32 (23\%) \\
 & dropout        & $\underset{[0.023,\,0.082]}{0.042}$                         & --- \\
 & l2             & $\underset{[9.29\text{e-}04,\,9.59\text{e-}03]}{5.43\text{e-}03}$ & --- \\
 & learning\_rate & $\underset{[2.58\text{e-}03,\,8.13\text{e-}03]}{5.69\text{e-}03}$ & --- \\
 & n\_layers      & $\underset{[3,\,4]}{3}$                                     & --- \\
 & neurons        & $\underset{[79,\,118]}{105}$                                & --- \\
 & optimizer      & ---                                                        & sgdwr (74\%), adam (13\%), adam\_amsgrad (10\%) \\
\midrule
\multirow{8}{*}{CoxTime}
 & activation     & ---                                                        & relu (97\%), tanh (3\%) \\
 & batch\_size    & ---                                                        & 128 (43\%), 64 (31\%), 32 (23\%) \\
 & dropout        & $\underset{[0.034,\,0.106]}{0.060}$                         & --- \\
 & l2             & $\underset{[7.64\text{e-}04,\,0.012]}{6.20\text{e-}03}$    & --- \\
 & learning\_rate & $\underset{[2.31\text{e-}03,\,7.80\text{e-}03]}{5.32\text{e-}03}$ & --- \\
 & n\_layers      & $\underset{[3,\,4]}{4}$                                     & --- \\
 & neurons        & $\underset{[90,\,119]}{109}$                                & --- \\
 & optimizer      & ---                                                        & sgdwr (73\%), adam\_amsgrad (17\%), rmsprop (7\%) \\
\bottomrule
\end{tabular}
\end{table}

\clearpage

\section{Simulated data\label{secA2}}

\subsection{Details of computation\label{subsecA21}}
\subsubsection{Data generation process 1\label{subsubsecA211}}
In this section, we introduce the calculation to obtain the theoretical survival values from the oracle model in the simulation data generation process 1.

To simulate survival times from a Weibull distribution, the following relationship can be used \citep{Bender2005}:
\begin{equation}
    T = H_{0}^{-1}[-\log(1-U)]\exp(\beta(X))
     \nonumber
\end{equation}
with $U\sim\mathcal{U}[0,1]$. In the case of a Weibull risk function, $\lambda>0$ is the scale parameter, $\kappa$ is the shape parameter, and the cumulative risk function is written as: 
\begin{equation}
     H_{0}(t) = \left(\frac{t}{\lambda}\right)^{\kappa}
      \nonumber
\end{equation}

The inverse of $H_{0}$ is then written:
\begin{equation}
    H^{-1}_{0}(u)=\lambda u^{1/\kappa} \nonumber
\end{equation}
We obtain: 

\begin{equation}
   \boxed{ T = \lambda [-\log(1-U)]^{1/\kappa}\exp(\beta X)}
   \nonumber
\end{equation}

 The basis risk function $h_{0}(t)$ is known and follows a certain probability distribution. In this case, we   use the Weibull distribution. The risk function is then written as:  

 \begin{equation}
   h(t|X) = \frac{\kappa}{\lambda}\left(\frac{t}{\lambda}\right)^{\kappa-1} \exp(\beta X) \nonumber
\end{equation}
 with $h_{0}(t)=\frac{\kappa}{\lambda}(\frac{t}{\lambda})^{\kappa-1}$ the risk at baseline and $\beta$ the coefficients vector.  We then obtain the survival function from the instantaneous risk :  
\begin{align}
    S(t|X) &= \exp(-\int_{0}^{t}h(s|X)ds  \nonumber \\
    &= \exp(-H_{0}(t)\exp(\beta X)). \nonumber
\end{align}

\subsubsection{Data generation process 2\label{subsubsecA212}}

In this section, we introduce the calculation to obtain the theoretical survival values from the oracle model in the simulation data generation process 2.

To simulate survival times from an AFT model and a random generator based on Friedman's function, the following relationship can be used \citep{Leemis1990}:
\begin{equation}
    T = \frac{H_{0}^{-1}[-\log(1-U)]}{\exp(m_{1}(X) + \tau\times m_{2}(X))}, \nonumber
\end{equation}
with $U\sim\mathcal{U}[0,1]$. In the case of a lognormal risk function, the cumulative risk function is written as :
\begin{equation}
    H_{0}(t) = -\log\left[1-\phi\left(\frac{\log(t)-\mu}{\sigma}\right)\right]. \nonumber
\end{equation}
The inverse of $H_{0}$ is then written:
\begin{equation}
    H_{0}^{-1}(u) = \exp(\sigma \phi^{-1}(1-exp(-u))+\mu). \nonumber
\end{equation}
We obtain:
\begin{equation}
   \boxed{ T = \frac{1}{\exp(m_{1}(X) + \tau\times m_{2}(X))}\exp(\sigma \phi^{-1}(U)+\mu)}. \nonumber
\end{equation}

 The basis risk function $h_{0}(t)$ is known and follows a certain probability distribution. In this case, we   use the log-normal distribution. The risk function is then written as:  
\begin{equation}
   h(t|X) = \exp(m_{1}(X) + \tau\times m_{2}(X))h_{0}(t \exp(m_{1}(X) + \tau\times m_{2}(X))), \nonumber
\end{equation}
with $h_{0}(t)$ the risk at baseline and $\beta$ the coefficients vector.  We then obtain the survival function from the instantaneous risk :  
\begin{align}
    S(t|X) &= \exp(-\int_{0}^{t}h(s|X)ds  \nonumber \\
    &= \exp(-H_{0}(t\exp(m_{1}(X) + \tau\times m_{2}(X))). \nonumber
\end{align}

\subsection{Kaplan Meier Curves for data generation process 1 and 2}

 We can see the Kaplan-Meier curves of the treatment group and the control group for one of the dataset in data generation process 1 and 2 in the following figures. 

\begin{figure*}[t]
\centerline{\includegraphics[width=0.6\textwidth]{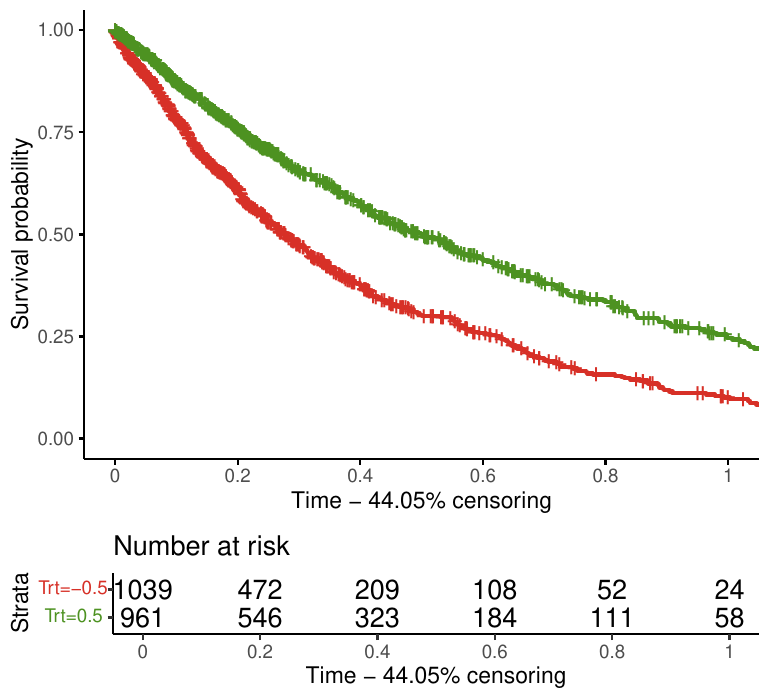}}
\label{fig:KMSim1}
\caption{Kaplan-Meier curve for one simulation set in data generation process 1, with $n=20000$.}
\end{figure*}

 \begin{figure*}[t]
\centering
\includegraphics[width=0.6\textwidth]{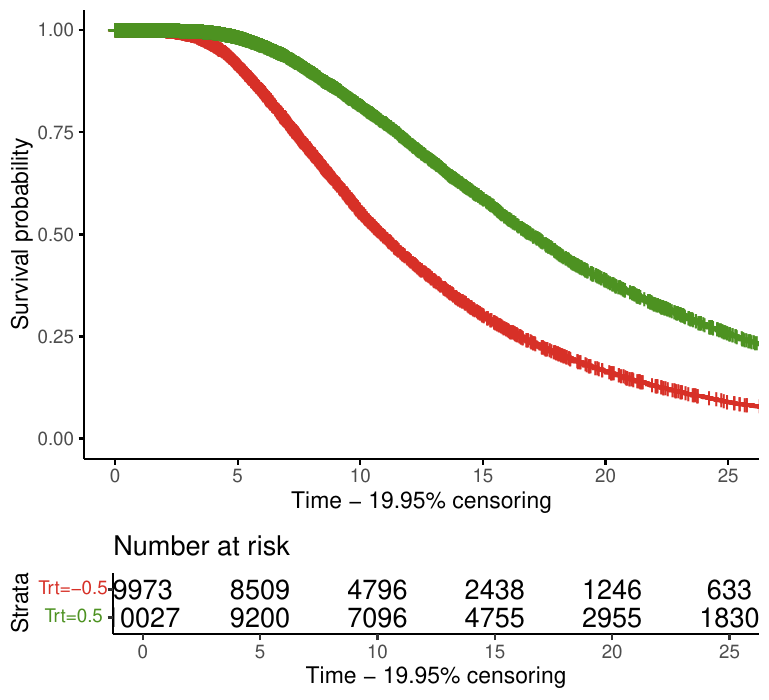}
\label{fig:KMSim2}
\caption{Kaplan-Meier curve for one simulation set in data generation process 2, with $n=20000$.}
\end{figure*}

\clearpage 
\subsection{Detailed results for data generation process 1\label{subsecA22}}

\begin{center}
\begin{table*}[ht]%
\caption{Average value at the first and third quartiles of time across the 100 simulation sets for data generation process 1.}\label{tab:Sett1P75C50}
{\tiny
\begin{tabular*}{\textwidth}{@{\extracolsep\fill}lllllll@{}}
\toprule
&\multicolumn{3}{@{}c}{\textbf{Q1}} & \multicolumn{3}{@{}c}{\textbf{Q3}} \\\cmidrule{2-4}\cmidrule{5-7}
\textbf{Model} &  \textbf{$\text{C}_{\text{benefit}}$}  & \textbf{$\text{E50}_{\text{benefit}}$}  & \textbf{$\text{ICI}_{\text{benefit}}$}  & \textbf{$\text{C}_{\text{benefit}}$}  & \textbf{$\text{E50}_{\text{benefit}}$}  & \textbf{$\text{ICI}_{\text{benefit}}$}   \\
\midrule
Oracle &  $\underset{(\pm0.022)}{0.604}$ &  &  & $\underset{(\pm0.024)}{0.504}$  &   &\\
ALASSO &   $\underset{(\pm0.021)}{0.501}$& $\underset{(\pm0.033)}{0.515}$ &  $\underset{(\pm0.030)}{0.515}$    & $\underset{(\pm0.020)}{0.503}$  &  $\underset{(\pm0.044)}{0.440}$  &  $\underset{(\pm0.042)}{0.439}$ \\
CoxCC &  $\underset{(\pm0.045)}{0.536}$ &  $\underset{(\pm0.137)}{0.395}$ & $\underset{(\pm0.134)}{0.394}$   & $\mathbf{\underset{(\pm0.039)}{0.538}}$ & $\underset{(\pm0.114)}{0.338}$ & $\underset{(\pm0.112)}{0.342}$ \\
CoxTime &$\underset{(\pm0.039)}{0.537}$  & $\mathbf{\underset{(\pm0.113)}{0.388}}$ & $\mathbf{\underset{(\pm0.127)}{0.388}}$   &$\underset{(\pm0.046)}{0.534}$   & $\mathbf{\underset{(\pm0.111)}{0.336}}$ & $\mathbf{\underset{(\pm0.110)}{0.339}}$\\
IF  &  $\mathbf{\underset{(\pm0.043)}{0.590}}$ & $\underset{(\pm0.051)}{0.499}$    &  $\underset{(\pm0.057)}{0.471}$  & $\underset{(\pm0.040)}{0.541}$  & $\underset{(\pm0.036)}{0.479}$ &  $\underset{(\pm0.031)}{0.466}$ \\
CSF & $\underset{(\pm0.042)}{0.511}$ & $\underset{(\pm0.045)}{0.064}$ & $\underset{(\pm0.040)}{0.073}$ & $\underset{(\pm0.036)}{0.509}$ & $\underset{(\pm0.041)}{0.064}$ & $\underset{(\pm0.037)}{0.072}$ \\
\bottomrule
\end{tabular*}}
\begin{tablenotes}%%[341pt]
\footnotesize
\item The value in brackets is the standard deviation across the 100 simulation sets. The highest $\text{C}_{\text{benefit}}$ values and lowest $\text{E50}_{\text{benefit}}$, $\text{ICI}_{\text{benefit}}$ and RMSE values are highlighted in bold. Patients are matched on predicted benefit. There is $50\%$ censoring.
\end{tablenotes}
\end{table*}
\end{center}

\begin{center}
\begin{table*}[ht]%
\caption{Average value at median time across the 100 simulation sets for data generation process 1.}
\label{tab:Sett1P5C20}
\resizebox{\linewidth}{!}{
\scriptsize
\begin{tabular*}{\textwidth}{@{\extracolsep\fill}lllllll@{}}
\toprule
&\multicolumn{3}{@{}c}{\textbf{Setting 1}} & \multicolumn{3}{@{}c}{\textbf{Setting 2}} \\\cmidrule{2-4}\cmidrule{5-7}
\textbf{Model} & \textbf{$\text{C}_{\text{benefit}}$}  & \textbf{$\text{E50}_{\text{benefit}}$}  & \textbf{$\text{ICI}_{\text{benefit}}$}   & \textbf{$\text{C}_{\text{benefit}}$}  & \textbf{$\text{E50}_{\text{benefit}}$}  & \textbf{$\text{ICI}_{\text{benefit}}$}   \\
\midrule
Oracle & $\underset{(\pm0.028)}{0.619}$   &  &  &      $\underset{(\pm0.028)}{0.559}$ &   &\\
ALASSO  & $\underset{(\pm0.032)}{0.501}$ &$\underset{(\pm0.035)}{0.684}$  & $\underset{(\pm0.034)}{0.682}$    &   $\underset{(\pm0.02)}{0.503}$ &   $\underset{(\pm0.042)}{0.442}$ & $\underset{(\pm0.04)}{0.443}$  \\
CoxCC & $\underset{(\pm0.046)}{0.568}$ &  $\underset{(\pm0.155)}{0.475}$ & $\underset{(\pm0.148)}{0.477}$   &     $\underset{(\pm0.036)}{0.528}$ &  $\underset{(\pm0.0099)}{0.350}$ &  $\underset{(\pm0.01)}{0.349}$ \\
CoxTime    & $\mathbf{\underset{(\pm0.045)}{0.575}}$ & $\underset{(\pm0.151)}{0.483}$ & $\underset{(\pm0.143)}{0.484}$  &     $\underset{(\pm0.034)}{0.531}$ & $\underset{(\pm0.084)}{0.360}$  & $\underset{(\pm0.084)}{0.357}$  \\
IF   & $\underset{(\pm0.053)}{0.562}$ & $\underset{(\pm0.027)}{0.694}$ & $\underset{(\pm0.027)}{0.679}$  &    $\mathbf{\underset{(\pm0.036)}{0.539}}$ & $\underset{(\pm0.032)}{0.479}$ & $\underset{(\pm0.037)}{0.488}$  \\
CSF & $\underset{(\pm0.024)}{0.519}$ & $\underset{(\pm0.036)}{\mathbf{0.07}}$ & $\underset{(\pm0.033)}{\mathbf{0.08}}$ & $\underset{(\pm0.024)}{0.510}$ & $\underset{(\pm0.034)}{\mathbf{0.063}}$ & $\underset{(\pm 0.032)}{\mathbf{0.071}}$ \\
\bottomrule
\end{tabular*}}
\begin{tablenotes}%%[341pt]
\footnotesize
\item The value in brackets is the standard deviation across the 100 simulation sets. The highest C values and lowest E for benefit and RMSE values are highlighted in bold. Patients are matched on predicted benefit. 
\item Setting 1: Matching on predicted benefit. There is $20\%$ censoring.
\item Setting 2: Matching on covariates.There is $50\%$ censoring.
\end{tablenotes}
\end{table*}
\end{center}

\begin{figure}[!ht]
  \centering
  \includegraphics[width=.8\linewidth]{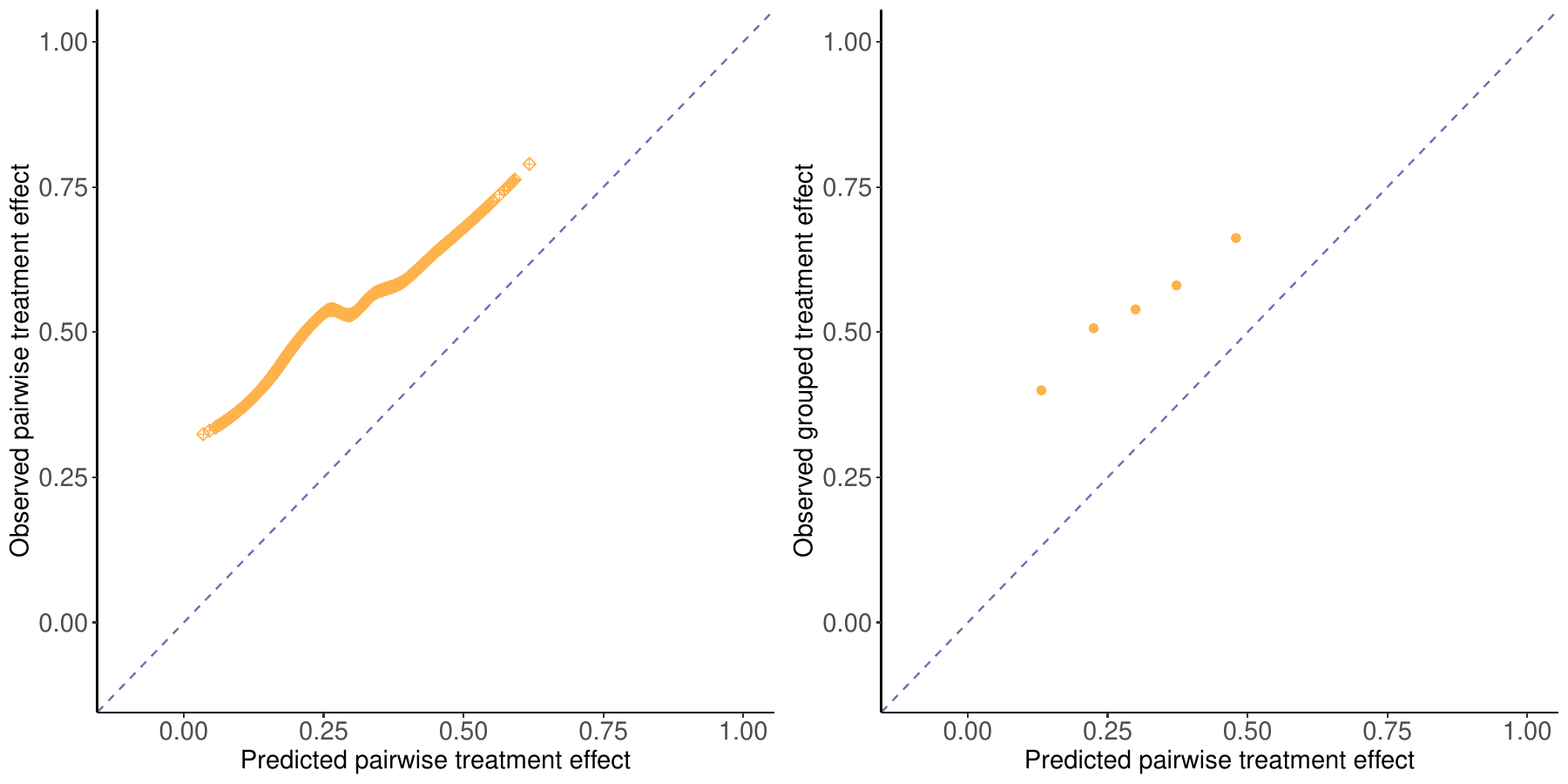}
  \caption{Calibration plot for one dataset in data generation process 1 and with CoxTime predictions. The left figure displays all the points, and the right figure shows quintiles.}
\end{figure}

\clearpage
\section{Variable Importance}

\begin{table}[htbp]
\centering
\caption{Top-5 predictive variables (treatment-effect modifiers) per model family, Breast. $P_{\text{top5}}$: proportion of bootstrap-fold replications in which the variable was ranked in the top 5. Mean rank is reported with the empirical 2.5\%--97.5\% interval of the rank distribution across replications.}
\label{tab:table18}
\begin{tabular}{llcc}
\toprule
Model & Variable & $P_{\text{top5}}$ & Mean rank [2.5\%, 97.5\%] \\
\midrule
\multirow{5}{*}{LASSO}
 & \texttt{X209398\_at}   & 0.34 & 23.0 [1.0, 95.0] \\
 & \texttt{X201005\_at}   & 0.20 & 24.8 [1.8, 96.0] \\
 & \texttt{X212095\_s\_at} & 0.19 & 29.1 [1.0, 94.5] \\
 & \texttt{X205683\_x\_at} & 0.16 & 29.4 [1.0, 88.7] \\
 & \texttt{X201552\_at}   & 0.15 & 27.4 [1.0, 98.1] \\
\midrule
\multirow{5}{*}{CoxCC}
 & \texttt{201551\_s\_at} & 0.04 & 2.5 [2.0, 3.8] \\
 & \texttt{201014\_s\_at} & 0.04 & 3.0 [1.2, 4.0] \\
 & \texttt{202555\_s\_at} & 0.03 & 1.7 [1.1, 2.0] \\
 & \texttt{201860\_s\_at} & 0.03 & 2.0 [1.1, 3.0] \\
 & \texttt{201116\_s\_at} & 0.03 & 2.3 [1.1, 3.0] \\
\midrule
\multirow{5}{*}{CoxTime}
 & \texttt{202016\_at}   & 0.03 & 1.7 [1.1, 2.0] \\
 & \texttt{201348\_at}   & 0.03 & 3.0 [1.1, 4.9] \\
 & \texttt{205541\_s\_at} & 0.03 & 3.2 [2.0, 6.6] \\
 & \texttt{201008\_s\_at} & 0.03 & 4.0 [1.2, 7.7] \\
 & \texttt{201650\_at}   & 0.03 & 5.8 [1.2, 14.2] \\
\midrule
\multirow{5}{*}{\shortstack[l]{Causal Survival\\Forest}}
 & \texttt{X211958\_at}   & 0.44 & 32.9 [1.0, 227.7] \\
 & \texttt{X208682\_s\_at} & 0.35 & 24.0 [1.0, 122.9] \\
 & \texttt{X210163\_at}   & 0.23 & 85.2 [1.0, 506.5] \\
 & \texttt{X204015\_s\_at} & 0.18 & 78.2 [1.0, 318.5] \\
 & \texttt{X202718\_at}   & 0.18 & 79.8 [1.0, 486.2] \\
\bottomrule
\end{tabular}
\end{table}

\begin{table}[htbp]
\centering
\caption{Top-5 predictive variables (treatment-effect modifiers) per model family, Trastuzumab. $P_{\text{top5}}$: proportion of bootstrap-fold replications in which the variable was ranked in the top 5. Mean rank is reported with the empirical 2.5\%--97.5\% interval of the rank distribution across replications.}
\label{tab:table19}
\begin{tabular}{llcc}
\toprule
Model & Variable & $P_{\text{top5}}$ & Mean rank [2.5\%, 97.5\%] \\
\midrule
\multirow{5}{*}{LASSO}
 & \texttt{DKFZP434A0131} & 0.40 & 14.5 [1.0, 54.5] \\
 & \texttt{THOP1}         & 0.29 & 21.5 [1.0, 83.9] \\
 & \texttt{C16orf14}      & 0.20 & 29.6 [1.0, 95.8] \\
 & \texttt{LOC442260}     & 0.16 & 27.7 [2.0, 93.0] \\
 & \texttt{METTL3}        & 0.15 & 25.0 [1.0, 90.2] \\
\midrule
\multirow{5}{*}{CoxCC}
 & \texttt{BTG2}    & 0.06 & 6.7 [2.2, 15.4] \\
 & \texttt{EMP2}    & 0.05 & 4.4 [2.0, 8.7] \\
 & \texttt{MPSIZ}   & 0.05 & 11.9 [1.0, 25.6] \\
 & \texttt{Kua.UEV} & 0.04 & 3.5 [3.0, 4.8] \\
 & \texttt{AK1}     & 0.04 & 8.1 [2.2, 16.6] \\
\midrule
\multirow{5}{*}{CoxTime}
 & \texttt{MPSIZ}  & 0.09 & 9.0 [2.3, 19.3] \\
 & \texttt{CASC3}  & 0.07 & 3.1 [1.0, 5.0] \\
 & \texttt{MAFK}   & 0.05 & 3.5 [1.1, 5.9] \\
 & \texttt{ENO1}   & 0.05 & 4.8 [2.1, 11.1] \\
 & \texttt{ADORA3} & 0.05 & 4.9 [1.0, 11.4] \\
\midrule
\multirow{5}{*}{\shortstack[l]{Causal Survival\\Forest}}
 & \texttt{CLIC1}   & 0.54 & 25.3 [1.0, 216.3] \\
 & \texttt{FAM83E}  & 0.47 & 17.5 [1.0, 92.7] \\
 & \texttt{GRB7}    & 0.18 & 62.2 [1.5, 279.0] \\
 & \texttt{DAD1}    & 0.14 & 91.8 [2.0, 336.7] \\
 & \texttt{B4GALT1} & 0.12 & 78.3 [2.0, 295.6] \\
\bottomrule
\end{tabular}
\end{table}

\begin{table}[htbp]
\centering
\caption{Top-5 predictive variables (treatment-effect modifiers) per model family, Avelumab. $P_{\text{top5}}$: proportion of bootstrap-fold replications in which the variable was ranked in the top 5. Mean rank is reported with the empirical 2.5\%--97.5\% interval of the rank distribution across replications.}
\label{tab:table20}
\begin{tabular}{llcc}
\toprule
Model & Variable & $P_{\text{top5}}$ & Mean rank [2.5\%, 97.5\%] \\
\midrule
\multirow{5}{*}{LASSO}
 & \texttt{MCDERMOTT\_ANGIO}                              & 0.92 & 1.7 [1.0, 5.0] \\
 & \texttt{B9991003\_c15\_Skin\_development}              & 0.38 & 5.6 [1.0, 16.0] \\
 & \texttt{B9991003\_c10\_Myogenesis}                     & 0.34 & 6.5 [2.0, 18.3] \\
 & \texttt{B9991003\_c17\_Glucocorticoid\_metabolic\_process} & 0.34 & 7.5 [1.0, 19.5] \\
 & \texttt{B9991003\_rc5\_Cell\_cell\_signaling}          & 0.31 & 7.5 [1.0, 21.0] \\
\midrule
\multirow{5}{*}{CoxCC}
 & \texttt{AGE}                                          & 0.93 & 2.0 [1.0, 7.0] \\
 & \texttt{B9991003\_c21\_Chromosome\_Y\_linked}         & 0.37 & 5.0 [1.0, 12.0] \\
 & \texttt{B9991003\_rc9\_Cell\_cycle}                   & 0.33 & 5.8 [1.0, 13.3] \\
 & \texttt{B9991003\_c23\_Cell\_adhesion\_at\_cell\_membrane} & 0.31 & 6.0 [1.0, 16.0] \\
 & \texttt{B9991003\_c18\_B\_cell\_activation}           & 0.25 & 7.2 [2.0, 20.2] \\
\midrule
\multirow{5}{*}{CoxTime}
 & \texttt{AGE}                                          & 0.91 & 2.3 [1.0, 9.2] \\
 & \texttt{B9991003\_c21\_Chromosome\_Y\_linked}         & 0.40 & 5.4 [1.0, 12.4] \\
 & \texttt{B9991003\_rc9\_Cell\_cycle}                   & 0.34 & 5.7 [1.0, 14.6] \\
 & \texttt{B9991003\_c23\_Cell\_adhesion\_at\_cell\_membrane} & 0.31 & 6.2 [1.6, 13.4] \\
 & \texttt{B9991003\_c20\_Interferon\_gamma\_response}   & 0.22 & 6.5 [1.1, 16.9] \\
\midrule
\multirow{5}{*}{IF}
 & \texttt{MCDERMOTT\_ANGIO}                             & 0.72 & 5.9 [1.0, 35.3] \\
 & \texttt{B9991003\_rc3\_Angiogenesis}                  & 0.49 & 11.7 [1.0, 50.0] \\
 & \texttt{AGE}                                          & 0.46 & 12.9 [1.0, 54.1] \\
 & \texttt{B9991003\_rc5\_Cell\_cell\_signaling}         & 0.42 & 12.3 [1.0, 41.0] \\
 & \texttt{B9991003\_c15\_Skin\_development}             & 0.33 & 17.6 [1.0, 52.5] \\
\midrule
\multirow{5}{*}{CSF}
 & \texttt{B9991003\_rc5\_Cell\_cell\_signaling}         & 0.62 & 5.6 [1.0, 19.0] \\
 & \texttt{B9991003\_c15\_Skin\_development}             & 0.46 & 9.2 [1.0, 26.5] \\
 & \texttt{MCDERMOTT\_ANGIO}                             & 0.46 & 9.3 [1.0, 30.0] \\
 & \texttt{B9991003\_c17\_Glucocorticoid\_metabolic\_process} & 0.35 & 10.2 [1.0, 31.0] \\
 & \texttt{B9991003\_c19\_Plasma\_cells}                 & 0.32 & 11.7 [1.0, 35.6] \\
\bottomrule
\end{tabular}
\end{table}

\end{document}